\documentclass[journal, a4paper,10pt]{IEEEtran}

\usepackage{amssymb}
\usepackage{amsmath}
\usepackage{graphicx}
\usepackage{multirow}
\usepackage{subcaption}
\usepackage{algorithm}
\usepackage{algpseudocode}
\usepackage{multicol}
\usepackage{xcolor}
\newcommand{\norm}[1]{\left\lVert#1\right\rVert}

\title{\huge{A Proactive Scalable Approach for Reliable Cluster Formation in Wireless Networks with D2D Offloading}} 
\author{Sanaa Sharafeddine and Omar Farhat\\ \textit{\{sanaa.sharafeddine, omar.farhat\}@lau.edu.lb}\\
Lebanese American University, Beirut, Lebanon}

\begin{document}
\maketitle
\begin{abstract}
With the current exponential growth in traffic and service demands, device-to-device (D2D) cooperation is identified as a major mechanism to enable 5G networks to effectively and efficiently augment network resources. The effectiveness of D2D cooperation depends on a wide range of decision making processes that include cluster formation, resource allocation, in addition to connection and mobility management. Irrespective of the D2D cooperation scenario whether in sensor, ad hoc, or cellular networks, the literature normally assumes that devices selected as relays or data sources are reliable; this means that they will maintain the connection till the communication session ends. Yet, this assumption is challenged in practice since devices' batteries can be depleted (e.g., sensors in an IoT network) and devices can move leading to connection termination (e.g., mobile users in a WiFi network or cars in a vehicular ad hoc network). To this end, we address the problem of reliable D2D cooperation in wireless networks by proposing a novel approach that is proactive by utilizing reliability metrics in the decision making process, and scalable by having low implementation complexity suitable for dense networks. These differentiating factors are shown to enhance the overall network reliability compared to standard techniques and to facilitate dynamic operation which is essential for practical implementation. Performance is evaluated using extensive simulations in addition to test bed experimental demonstration in order to quantify gains and extract insights on a range of existing design tradeoffs.
\end{abstract}

\section{Introduction}
\label{sec:introduction}
Device-to-device (D2D) communication is expected to play a key role in 5G systems to provision ultra dense networks with improved performance, reduced latency,  and lower energy consumption. Forming reliable D2D communications is yet another challenge that is foreseen to realize ultra-reliable low latency communications. Cooperation among devices has been shown to be highly effective in enhancing the performance of infrastructure-based wireless networks for a wide range of use cases; devices in the context of this work can be sensor nodes in an Internet of Things~(IoT) network, mobile users in a cellular or WiFi network, or cars in a vehicular ad hoc network~(VANET)~\cite{BZ16,GJ16, TUY14, CGS13, FDM12}. Facilitating mobile-to-mobile, sensor-to-sensor or car-to-car (all denoted as device-to-device~(D2D) in the sequel) cooperation can lead to traffic offloading, throughput enhancement, energy consumption savings, coverage extension, and/or cost reduction, e.g., see~\cite{Detti2015, LJK15, Mumtaza2014, Jahed2012} and the references therein.

A major challenge that hinders reaping the benefits of cooperation in wireless networks is reliability and robustness against device-level dynamics as channel conditions vary, devices move, and devices' data traffic and energy/processing capabilities change over time. In infrastructure-based cooperative wireless networks, an essential element is the grouping of nodes into clusters whereby cluster head devices are selected intelligently to act as relays between infrastructure nodes (e.g., cellular base stations or WiFi access points) and other devices; the data communication within any cluster can be disrupted upon the loss of its cluster head device, due to battery drain, device mobility, device malfunctioning, or even a malicious security attack. To this end, in this work we focus on the design of a proactive scalable approach for reliable cluster formation in cooperative wireless networks while achieving target performance requirements and having low complexity to be practically applicable to high dense scenarios.


\subsection{Related Work}
\label{sec:related}

Even though there is rich literature on clustering techniques and solutions for cooperative wireless networks (e.g., see the surveys~\cite{SRM16, ACF17} and the references therein), there is relatively less progress on addressing challenges related to fault tolerance and reliability. Failure recovery and survivability have been addressed in the context of wireless sensor and ad hoc network scenarios (e.g., see~\cite{Haider2015, Younisa2014, Kofahi2010}). In sensor networks, failure recovery typically aims at maintaining connectivity as sensor devices die due to energy depletion, whereas in ad hoc networks a key aim is to adapt multihop routing in order to maintain network level connectivity as nodes move around. The related literature can in general be divided into reactive (post clustering) and proactive (pre clustering) reliability enhancement techniques.

Reactive techniques aim at either minimizing failures or recovering from failure events in a dynamic fashion after cluster formation. The authors in~\cite{GZH12} present an RSSI (received signal strength indicator) based approach for cluster formation in sensor networks whereby cluster head selection takes into account signal strength level and energy budget among neighbor nodes with the option of dynamically replacing cluster heads based on given performance metrics. The authors in~\cite{SY15} present a mechanism for dynamic cluster head re-election by executing an update algorithm on a periodic basis using pre-configured parameters. The authors in~\cite{AKJ15} address fault tolerance in wireless sensor network scenarios using a distributed real-time recovery algorithm based on periodic protocol exchange messages to identify failures and deal with them. In~\cite{SJF17}, the authors presented several low complexity reactive algorithms to deal with mobility in content distribution networks with D2D cooperation while taking into account three classes of events: an existing device leaves the network which is highly challenging when the device is a cluster head, an existing device moves locally within the network, and a new incoming device joins the network.

On the other hand, proactive techniques take preventive measures as part of the cluster formation process, either by considering device reliability metrics when electing cluster heads (also called group owners) or by identifying and assigning backup cluster heads to deal with failure events. For example, the authors in~\cite{CYF15} present an interesting approach for group re-formation in WiFi-Direct D2D networks that includes the election of an emergency group owner serving as a backup cluster head and the configuration of dormant backend links for fast group establishment after failure events. The authors in~\cite{PJ17, MMZ15,SRV14, JZZ16} present robust clustering algorithms for different network scenarios based on backup (also called secondary or redundant) cluster head selection taking into account metrics such as node degree and energy consumption. A hierarchical fuzzy logic based approach was developed in~\cite{ND16} whereby cluster heads connect to a super cluster head node acting as relay to a mobile base station. Short range D2D wireless technologies such as WiFi-Direct and Bluetooth do not include intelligence for group owner (cluster head) election as part of their standards. This triggered research to devise optimized mechanisms for cluster head election, either based on rotation among devices to distribute load over time such as the LEACH~(low-energy adaptive clustering hierarchy) protocol~\cite{HCB02} or based on biased selection taking into account performance metrics such as centrality of location or availability of energy budget~\cite{CKF17, JFA16,GZH12, MCB14}. For example, the LEACH protocol divides time into rounds whereby in each round a new random sensor node is assigned as cluster head to evenly distribute the energy load among all sensors over time.

\subsection{Contributions}

In this paper, we present a novel proactive approach for reliable cluster formation among devices in infrastructure-based wireless network with D2D cooperation. The key differentiating factors of the proposed approach are its flexibility and scalability. The approach is flexible in proactively capturing different reliability metrics as part of a generic cost function and is flexible in its applicability to different wireless network scenarios including sensors, IoT devices, and mobile users. On the other hand, the approach is scalable to network scenarios with high density of devices as it has a relatively fast execution time. It is important to highlight that the proposed proactive approach can complement a wide range of existing state-of-the-art reactive techniques to further enhance the level of reliability. Thus, our proactive approach runs before initiating a given service to decide on cluster heads that will be managing D2D communications. During service operation and in the case of an unexpected cluster head failure, a complementary reactive approach has to be implemented. The reader is referred to Section~\ref{sec:related} for a list of existing reactive approaches that can be incorporated.

In terms of solution methodology, we formulate the problem as an integer linear program and generate optimal results for small scale scenarios. For large scale scenarios, we utilize a fast and effective algorithm based on the notion of electrostatic forces; we demonstrate its effectiveness in achieving close-to-optimal results and its scalability by generating results for high dense environments with multiple access points. Finally, we extend the contributions to experimental evaluation using test bed implementation in order to demonstrate the algorithm's feasibility and effectiveness under realistic operational conditions. It is important to note that very few studies in the literature include practical test bed implementation due to the challenges in integrating intelligence into devices especially when dealing with off the shelf smartphones; for example, see~\cite{LSY16, JSM16} for WiFi-Direct ad hoc network formation among Android phones, however, without dealing with intelligence related to cluster head election. Moreover, the authors in~\cite{SY15} present Android based implementation to demonstrate the practical implementation of their group creation and inter-communication approach using a network scenario composed of five devices.

In brief, the contributions of this work are multi-fold and can be summarized as follows: 1) A novel cluster formation mechanism in wireless networks with device-to-device cooperation is proposed to optimize network reliability; 2) A heuristic approach with close-to-optimal performance and high efficiency is proposed for networks with high density of devices such as IoT and WSN scenarios; 3) An experimental testbed is developed to demonstrate the practicality and effectiveness of the proposed algorithms in real scenarios.

\subsection{Organization}
Section~\ref{sec:model} presents the system model including key assumptions and metrics. Section~\ref{sec:prob} includes the optimization problem formulation with explanation of the objective function, constraints, and problem complexity. Section~\ref{sec:method} presents the solution methodology including the details of the utilized algorithm; results and analysis are then summarized in Section~\ref{sec:results} for a wide range of scenarios with various network and design parameters, including experimental evaluation. Finally, conclusions are drawn in Section~\ref{sec:conclusions}.

\section{System model}
\label{sec:model}

\begin{figure}[h]
	\centering
	\includegraphics[width=\columnwidth]{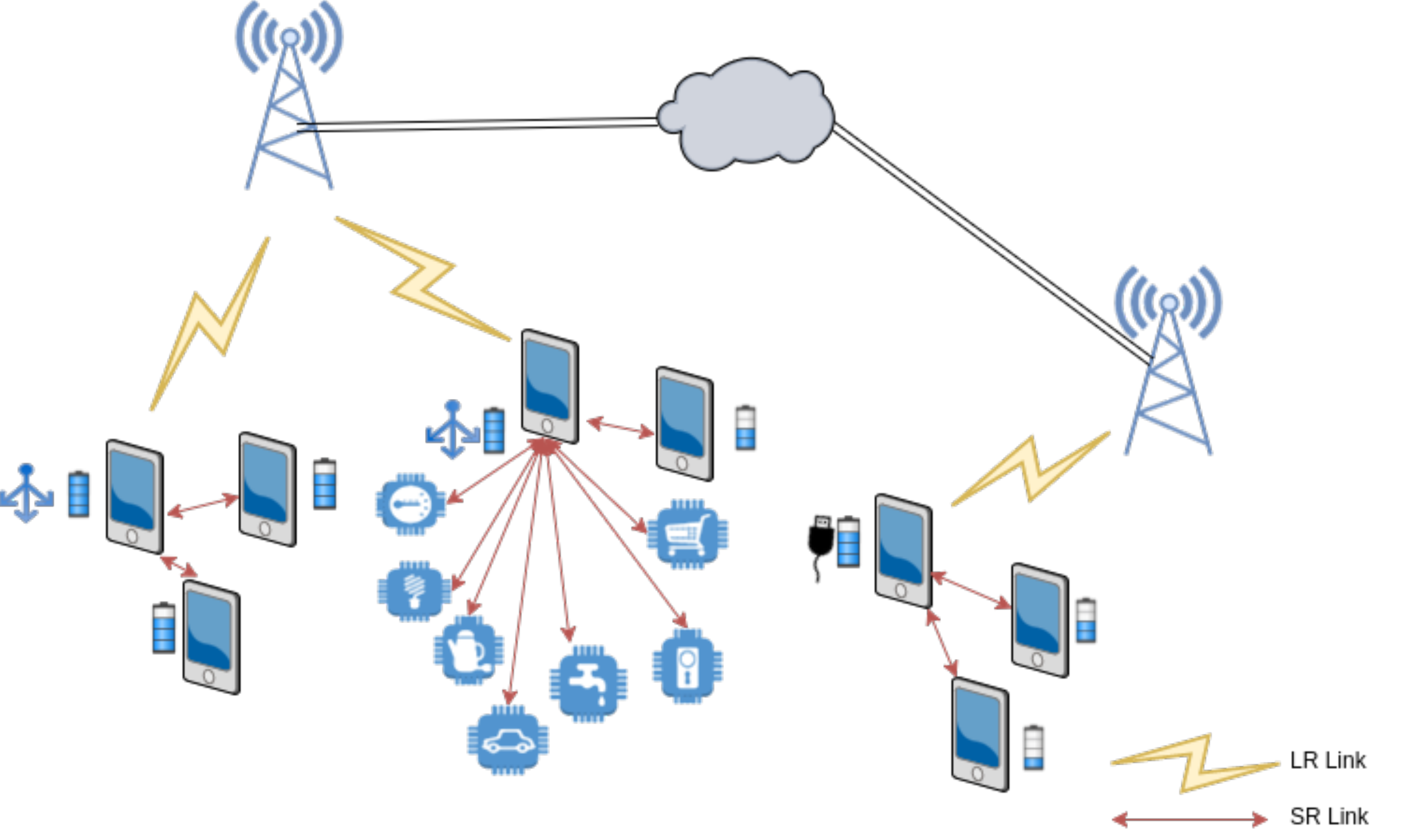}
	\caption{Example system model.}
	\label{fig:system_model}
\end{figure}

We consider a wireless network scenario composed of $N$ devices (sensors, IoT nodes, or mobile users) and $M$ WiFi access points~(APs) or cellular base stations in a given geographic area. The devices are equipped with multiple wireless interfaces supporting long range~(LR) connectivity to the APs and short range~(SR) connectivity to other device over direct D2D links (e.g., WiFi-Direct, LTE-Direct, or Bluetooth). It is worth noting here that such setup may lead to notable energy savings since communications within the cluster take place over short range connections that are more energy efficient than long range links~\cite{Jahed2012, balasubramanian2009energy, 6952687}. Moreover, we assume centralized intelligence in the network side to facilitate D2D cooperation whereby the devices are divided into clusters with one device per cluster selected as a cluster head. The cluster head acts then as a relay for data download between the APs and other devices in its cluster. Even though this is a standard system model, the novelty and differentiating factor of our work is the focus on reliability in the cluster formation process in order to minimize failure costs and enhance overall network performance.  Fig.~\ref{fig:system_model} presents a general schematic of the adopted system model and highlights that each device has certain characteristics reflected via battery status, availability of multiple interface, etc.

We denote the reliability of a given device $i$ as $\Gamma_i$, where $\Gamma_i$ is a value between $0$ and $1$ and is a function of  multiple factors including battery life denoted as $E_i$ and device rating denoted as $\gamma_i$. For the energy metric, we define a percentage threshold $\beta$ below which the device cannot cooperate due to limited budget and, thus, is marked non-reliable. On the other hand, for the device rating $\gamma_i$, we assume it is a value between~0 and~1 based on learning from historical data on the device's performance with respect to reliable cooperation, for example, based on the device's mobility patterns or D2D data sharing success rate. Connection failure due to mobility is a minor concern in limited mobility environments such as static WSN or IoT. In the case of more dynamic environments, devices  with unexpected mobility pattern that disturbs existing links among neighboring devices would receive a lower $\gamma_i$ to avoid selecting them as cluster heads in future D2D communications. We use the following expression to model the device's reliability, which gets non-zero value as long as $\gamma_i$ is non-zero and the actual battery level is above the threshold $\beta E$ where $E$ represents the full battery capacity; the $E - \beta E$ factor in the denominator is used for normalization purposes to constrain $\Gamma_i$ between 0 and 1:
\begin{equation}
	\label{eq:reliability}
	\Gamma_i = \gamma_i \min \left(\frac{E_i - \beta E}{E - \beta E}, 0\right)
\end{equation}

Let $\mathcal{D} = \{d_1, d_2, \cdots, d_N\}$ denote the set of devices and $\mathcal{S} = \{s_1, s_2, \cdots, s_M\}$ denote the set of available access points. The system can then be modeled as a graph $G = (V, W)$, where $V = \mathcal{D} \cup \mathcal{S}$ and $W$ the set of edges containing edge $(u,v)$ if and only if two nodes $u$ and $v$ are within radio range with respect to each other.
Moreover, we define the set $\mathcal{C} \subset \mathcal{D}$ composed of cluster head devices selected  to relay downloaded data to other devices in the set $\mathcal{D} - \mathcal{C}$ using SR communications links. Thus, data can be received either over a LR link directly from the AP or SR link from a neighboring node in $\mathcal{C}$.

\section{Problem Formulation}
\label{sec:prob}

\begin{table}[]
\centering
\caption{Notations used in problem formulation}
\label{params-variables-table}
\begin{tabular}{|p{1.4cm} |p{6.6cm}|}
\hline
Notation                & Description                            \\ \hline
$N$                                            & Number of devices                                          \\\hline
$M$                                            & Number of WiFi access points or cellular base stations                               \\\hline
$\mathcal{A}$                                  & $N\times N$ association matrix that indicates D2D connections among devices over SR                 \\\hline
$a_{ij}$ & binary variable that indicates if device $i$ has device $j$ as cluster head\\\hline
$\mathcal{B}$                                  & $M\times N$ association matrix between APs and devices over LR         \\\hline
$b_{mi}$ & binary variable that indicates if device $i$ is associated to AP $m$ \\\hline
$C_i$                                          & Cost function that models damage caused to network upon the failure of device $i$ being assigned as cluster head    \\\hline
$P_{l, mi}$                                    & Power received by device $i$ from AP $m$ over LR       \\\hline
$P_{s, ij}$                                    & Power received by device $j$ from device $i$ over SR D2D   \\\hline
$\sigma^2$                                     & Total noise level over any connection                  \\\hline
$\Gamma_i$                                     & Reliability metric of a given device $i$                      \\\hline
$\rho$                                         & Tradeoff between network reliability and performance   \\\hline
$\Delta_{\mathrm{LR}}$ & Maximum number of devices connected to one AP\\\hline
$\Delta_{\mathrm{SR}}$ & Maximum number of devices connected to one cluster head \\\hline
$\theta$                                       & Outage probability set by network operator         \\ \hline
$SNR_{LR}$ & minimum SNR threshold on LR\\\hline
$SNR_{SR}$ & minimum SNR threshold on SR \\\hline
\end{tabular}
\end{table}

We formulate the problem of reliable cluster formation with performance constraints as an integer linear programming problem. We define $N \times N$ association matrix $\mathcal{A}$ such that $a_{ij}$ is a binary value that determines if device $i$ is sharing data with device $j$ over a SR connection. Moreover, we define an $M \times N$ association matrix $\mathcal{B}$ such that $b_{mi}$ is a binary value that determines if AP $m$ is transmitting data directly to device $i$ over a LR connection. We denote by $P_{l,mi}$ the power received by device~$i$ from AP~$m$ over a LR connection and by $P_{s,ij}$ the power received by device $j$ from neighboring device $i$ over a SR D2D connection.
	
The goal is to select intelligently a set of devices $\mathcal{C}$ as cluster heads while jointly optimizing network reliability and performance. We optimize performance by maximizing the received power level at all devices irrespective if over LR connections from an AP or over SR connection from another nearby device. The received power level can then be mapped to data rate using $R = W\log_2(1 + \mathrm{SNR})$ where $W$ is bandwidth, $\mathrm{SNR} = {P_{\mathrm{received}}}/{\sigma^2}$, $P_{\mathrm{received}}$ is received power level either on SR or LR connection, and $\sigma^2$ represents total noise level. In addition to improving performance in terms of increased bit rate, the clustering approach also leads to energy savings while transmitting data among devices. As the received power level is maximized, devices within a cluster are selected in close proximity to each other and can transmit their data with reduced power leading to major energy reduction gains. 

%
%
%

In order to model network reliability, we define a generic cost function $C_i$ that models the damage caused to the network when device $i$ acting as cluster head fails while sharing content with other nearby devices in its cluster. We model the failure cost of device $i$ as a function of its individual reliability metric $\Gamma_i$ defined in~(\ref{eq:reliability}) in addition to the number of devices it serves on the SR, i.e., number of devices in its cluster. Devices with relatively lower battery levels are allocated a reduced reliability metric and, thus, the solution opts to eliminate them from being selected as cluster heads. In addition, when the number of devices served by a cluster head increases, its failure cost grows as it impacts a larger pool of devices. The failure cost $C_i$ can be expressed as follows:
   	\begin{equation}
	\label{cost}
	C_i = (1 - \Gamma_i) \times \sum_{j = 1, j \neq i}^{N} a_{ij}
\end{equation}

The optimization problem can then be formulated as follows:

\begin{align}
	\label{obj_function}
	&\text{minimize } \rho \sum_{i = 1}^{N} C_i - \sum_{m = 1}^{M} \sum_{i = 1}^{N} b_{mi}  P_{l, mi} - \sum_{i = 1}^{N}\sum_{j = 1, j\neq i}^{N} a_{ij} P_{s, ij} 
\end{align}

 subject to \\
 \begin{align}
  	\label{constraint_1}
  	& a_{ij} \leq \sum_{m = 1}^{M} b_{mi}   &i, j = 1 ,..., N \\
	\label{constraint_2}
	 &\sum_{m = 1}^{M} b_{mi} + \sum_{j = 1, j \neq i}^{N} a_{ji} \leq 1 		  &i = 1, ..., N    \\
	\label{constraint_3}
	 &\sum_{i = 1}^{N} b_{mi} \leq \Delta_{\mathrm{LR}}                             		  & m = 1, ..., M    \\
	\label{constraint_4}
	 &\sum_{j = 1}^{N} a_{ij} \leq \Delta_{\mathrm{SR}}                 &i = 1, ..., N    \\
	\label{constraint_5}
	&\sum_{m=1}^{M} \sum_{i=1}^{N} b_{mi} + \sum_{i = 1}^{N} \sum_{j = 1}^{N} a_{ij} \geq (1 - \theta)N & \\
	\label{constraint_6}
	 &\frac{P_{l, mi}}{\sigma^2} \geq b_{mi}  \mathrm{SNR_{LR}}                 & \forall m,i \\
	\label{constraint_7}
	&\frac{P_{s, ij}}{\sigma^2} \geq  a_{ij} \mathrm{SNR_{SR}}                & \forall i,j 
\end{align}

The objective function in~(\ref{obj_function}) balances a trade-off between maximizing the overall received power levels over all LR and SR connections and minimizing the failure cost  for each cluster head device in $\mathcal{C}$. The parameter $\rho$ is introduced to balance the tradeoff between network reliability and performance; for instance, if $\rho$ is large, then the solution would favor producing a more reliable network at the cost of overall performance and vice-versa.

The first constraint in~(\ref{constraint_1}) specifies that device $i$ can transmit data to device $j$ over a SR connection only when it is receiving its content from an AP over a LR connection. We limit the level of cooperation to two hops for practical feasibility reasons, especially in use cases where devices are low end IoT sensors. The second constraint in~(\ref{constraint_2}) limits the active reception data links of any device to one; it is satisfied if and only if a device is receiving data over either a LR link or SR link but not both simultaneously. Constraints represented in~(\ref{constraint_3}) and~(\ref{constraint_4}) limit the number of devices connected to the AP and to a cluster head to $\Delta_{\mathrm{LR}}$ and $\Delta_{\mathrm{SR}}$, respectively; these constraints are added to capture the limit on the number of transmission channels per AP or device. The constraint in~(\ref{constraint_5}) bounds the outage probability in the network; the number of served devices is set to a minimum of $(1- \theta)N$ where $\theta$ represents the outage probability set by the network operator. Finally, the constraints in~(\ref{constraint_6}) and~(\ref{constraint_7}) require the received SNR per device to be above a minimum threshold on both LR ($\mathrm{SNR_{LR}}$) and SR ($\mathrm{SNR_{SR}}$) connections in order to maintain a target level of performance guarantee. The values of the various thresholds and parameters can be configured depending on the application scenario and requirements.

The formulated problem is an integer linear programming~(ILP) problem and can be mapped to the K-medoids problem, which is $\mathcal{NP}$-hard~\cite{Christos0210040, Nimrod0213014}. The K-medoids problem entails selecting $K$ centroids among  devices such that the aggregated distance between the centroids and devices is minimized. Our problem can actually be reduced to K-medoids by relaxing problem constraints. For instance, in the case of a single AP, we set the reliability of all devices to $1$ and set system parameters $\rho, \theta$, $\Delta_{\mathrm{LR}}$, $\Delta_{\mathrm{SR}}$ to $0$, $0$, $K$, and $N$, respectively. Consequently, we are only looking for centroids that minimize the distance between them and other nodes, which maps to the K-medoids problem. In addition, our problem becomes harder with multiple APs since centroid selection is also affected by the deployed position of the APs, as we attempt to also minimize the distance between each AP and the centroids (cluster heads) that it serves. To this end, we generate in the results section optimal solutions only for relatively small scale scenarios. In addition, we propose in the following section an efficient proactive heuristic algorithm for reliable cluster formation that is scalable to large scenarios with close-to-optimal performance.

\section{Solution Methodology and Algorithms}
\label{sec:method}


The problem formulation in~(\ref{obj_function}) can be solved for relatively small scale network scenarios to generate optimal reliable clustering results. For large scale network scenarios, we propose the utilization of a clustering algorithm based on the notion of electrostatic forces~\cite{Force}. The authors in~\cite{Force,Force1} demonstrated the validity and effectiveness of an approach based on electrostatic forces for applications that include general data clustering and tumor detection in images. In our work, we capitalize on this approach and extend it to address the problem of reliable cluster formation in cooperative wireless networks.

The algorithm starts by randomly scattering $K$ virtual centroids on the given area and then iteratively utilizes the law of electrostatics to determine optimized positions for the centroids; we describe these centroids as ``virtual'' since their locations need not overlap with existing devices in the network which requires an extension phase of the algorithm implementation to map each ``virtual'' centroid to one of the existing devices that would serve as a cluster head. In the algorithm, each device is assigned a negative fixed charge while virtual centroids are assigned dynamic positive charges; hence, the force among centroids is repulsive, while the force between centroids and devices is attractive. In the context of our problem, we let the charge of each device  be a function of its own reliability function as follows:
\begin{equation}
	\label{eq:device_charge}
	Q_i = -\Gamma_i.
\end{equation}

This allows devices to be clustered based on their reliability and relative positions. As for identifying the best centroid locations, we set the charge of a given centroid $k$ to be inversely proportional to the number of devices $N_k$ associated to it. Thus, the charge of centroid $k$ changes in every iteration as follows:
\begin{equation}
	\label{eq:centroid_charge}
	Q_k = \frac{\lambda}{N_k + 1}.
\end{equation}
\noindent where $\lambda$ is a pre-set parameter with value between $0$ and $1$; it can be configured to control the distance between the virtual centroid locations in the final solution.

In the network, there exist two main types of forces, namely, repulsion and attraction. Virtual centroids repel each other because they carry like charges while a centroid and a device attract each other because they carry unlike charges. Based on Coulomb's law, the force experienced by centroid $k$ in the vicinity of centroid $j$ can be calculated as follows using a unit vector to determine the force direction:
\begin{equation}
	\label{eq:force_centroids}
	\vec{F_{jk}} = \kappa \frac{Q_kQ_j}{d_{k,j}^2}\times \frac{\vec{c_k} - \vec{c_j}}{\norm{\vec{c_k} - \vec{c_j}}},
\end{equation}
\noindent where $\kappa$ is Coulomb's constant, $d_{k,j}$ is the Euclidean distance between $k$ and $j$, and $\vec{c_k}$ and $\vec{c_j}$ represent the centroid coordinate vectors, respectively.

The electrostatic force of attraction exerted by device $j$ on centroid $k$ is similarly calculated as follows: 	
\begin{equation}
	\label{eq:force_device}
	\vec{F_{jk}} =  \kappa \frac{Q_kQ_j}{d_{k,j}^2}\times \frac{\vec{c_k} - \vec{p_j}}{\norm{\vec{c_k} - \vec{p_j}}}.
\end{equation}
\noindent where $\vec{p_j}$ is the device coordinate vector.

Consequently, an electric field is formed among the nodes causing centroids to repel from each other and attract to devices. Hence, centroids move until electrostatic equilibrium is reached where the sum of forces is balanced and centroids are fixed. We denote the total force exerted on centroid $k$ as $\vec{F_k}$ and calculate it as the summation of all repulsion and attraction forces as follows:
\begin{equation}
\begin{split}
    \label{eq:sumForces}
    \vec{F_{k}} & = \sum_{j\neq k}{\vec{F_{jk}}}\\
    & = \sum_{j\neq k, j\in\mathcal{C}}{\kappa \frac{Q_kQ_j}{d_{k,j}^2}\times \frac{\vec{c_k} - \vec{c_j}}{\norm{\vec{c_k} - \vec{c_j}}}} + \sum_{j\in\mathcal{D}}{\kappa \frac{Q_kQ_j}{d_{k,j}^2} \times \frac{\vec{c_k} - \vec{p_j}}{\norm{\vec{c_k}-\vec{p_j}}}}.
\end{split}
\end{equation}

The direction in which centroids move is determined according to the sum of forces exerted on each centroid with a pre-configured step size $\eta$. Fixing the step size, we can calculate the centroid's new position as follows:

\begin{equation}
	\label{eq:force_step}
	\vec{c_k}^{\tau + 1} = \vec{c_k}^{\tau} + \eta\frac{\vec{F_k}}{\norm{\vec{F_k}}},
\end{equation}

\noindent where $\tau$ represents the iteration number reflecting the algorithm's execution over time. The algorithm runs over multiple iterations till it reaches a stable state, whereby the virtual centroids positions vary only locally within a circle of small radius. Therefore, the algorithm's implementation keeps track of the position variation increments per centroid to decide when to stop and generate a solution.

The presented solution approach is denoted as $\mathrm{RForce}$ and is divided into three main phases: Phase~1, summarized in Algorithm~1 and Algorithm~2, optimizes the locations of virtual centroids using proposed approach based on electrostatic forces; Phase~2, summarized in Algorithm~3, maps the virtual centroids to existing devices that will act as cluster heads; and Phase~3, summarized in Algorithm~4, associates cluster heads with APs in an optimized way that enhances the download bit rate while satisfying the given constraints.



\begin{algorithm}[t]
	\caption{\textsc{RForce - Phase 1}}
	\label{RForce-P1}
	\begin{algorithmic}[1]	
	\Procedure{RForce}{$\mathcal{C}$, $\mathcal{D}$}\\ 
	{\color{darkgray}\textsc{Input:} Set of centroids $\mathcal{C}$ and set of devices $\mathcal{D}$\\
    \textsc{Output:} Association matrix $\mathcal{A}$ and optimized location of all centroids }
		\While {\textsc{Not\_Stable}($\mathcal{C}$)}
		\For{each centroid $c_k \in \mathcal{C}$}
		\State $c_k$.degree $\gets 0$
		\EndFor
		\State $\mathcal{A} \gets \textsc{Associate\_Centroids}(\mathcal{C}, \mathcal{D})$
		\State $\mathcal{F} \gets \emptyset$
		\For{each centroid $c_k \in \mathcal{C}$}
		\For{each node $v \in \mathcal{C} \cup \mathcal{A}_k$}
		\State $\mathcal{F}_k \gets \mathcal{F}_k + \textsc{Force}(c_k, v)$ 
		\EndFor
		\EndFor
		\State $\textsc{Move\_Centroids}(\mathcal{C}, \mathcal{F})$ 
		\EndWhile
		\EndProcedure
\\
		\Procedure{Not\_Stable}{$\mathcal{C}$}
		\\
	{\color{darkgray}\textsc{Input:} Set of centroids $\mathcal{C}$\\
    \textsc{Output:} TRUE if at least one centroid $c_k$ changed its position, FALSE otherwise}
		\For{each centroid $c_k \in C$}
		\If{$c_k.x \neq c_k.x'$ OR $c_k.y \neq c_k.y'$}
		\State \Return $TRUE$
		\EndIf
		\EndFor
		\State \Return $FALSE$
		\EndProcedure
	\\	
	\Procedure{Move\_Centroid}{$\mathcal{C}$, $\mathcal{F}$}\\
		{\color{darkgray}\textsc{Input:} Set of centroids $\mathcal{C}$, and corresponding force $\mathcal{F}$ exerted on them\\
		\textsc{Output:} Updated coordinates of each centroid}
		\For{each centroid $c_k \in \mathcal{C}$}
		\State $c_k.x' = c_k.x$
		\State $c_k.y' = c_k.y$
		\State $c_k.x' = c_k.x + \eta \frac{\mathcal{F}_k.x}{\norm{\mathcal{F}_k}}$\vspace{0.1cm}
		\State $c_k.y' = c_k.y + \eta \frac{\mathcal{F}_k.y}{\norm{\mathcal{F}_k}}$
		\EndFor
		\EndProcedure
	\algstore{rforce-phase1}
	\end{algorithmic}
\end{algorithm}

\begin{algorithm}[th!]
    \caption{\textsc{RForce - Phase 1 \textit{cont'd} \\ \textsc{Associate\_Centroids} and \textsc{Best\_Centroid} Methods}}
	\label{RForce-P1B}
	\begin{algorithmic}[1]
	\algrestore{rforce-phase1}		
		\Procedure{Associate\_Centroids}{$\mathcal{C}$, $\mathcal{D}$}\\
		{\color{darkgray}\textsc{Input:} Set of centroids $\mathcal{C}$ and set of devices $\mathcal{D}$\\
    \textsc{Output:} Association matrix $\mathcal{A}$ that indicates devices mapped to each centroid}
        \State $\mathcal{A} \gets \emptyset$
		\For{each device $d \in \mathcal{D}$}
		\State $c_k \gets \textsc{Best\_Centroid}(\mathcal{C}, d)$
		\State $c_k$.degree $\gets c_k$.degree + 1;
		\State $\mathcal{A}_k \gets \mathcal{A}_k \cup d$
		\EndFor
		\State \Return $\mathcal{A}$
		\EndProcedure \\
		
		\Procedure{Best\_Centroid}{$\mathcal{C}$, $d$}
		\\
		{\color{darkgray}\textsc{Input:} Set of centroids $\mathcal{C}$ and certain device $d$\\
		\textsc{Output:} Best centroid for device $d$}
		\State best\_centroid $\gets \emptyset$
		\State min\_distance $\gets \infty$
		\For{each centroid $c_k \in \mathcal{C}$}
		\State distance $\gets$ \textsc{Distance}($c_k.x$, $c_k.y$, $d.x$, $d.y$)
		\If{$c_k$.degree $< \Delta_{\mathrm{SR}}$ AND distance $<$ min\_distance}
		\State min\_distance $\gets$ distance
		\State best\_centroid $\gets$ $c_k$
		\EndIf
		\EndFor
		\State \Return best\_centroid
		\EndProcedure 
	\end{algorithmic}
\end{algorithm}		

\begin{algorithm}[th!]
	\caption{\textsc{RForce - Phase 2}}
	\label{RForce-P2}
	\begin{algorithmic}[1]	
		\Procedure{Map\_Centroids}{$\mathcal{C}$, $\mathcal{D}$, $\mathcal{A}$}
		\\
		{\color{darkgray}\textsc{Input:} Set of centroids $\mathcal{C}$, set of devices $\mathcal{D}$, and association matrix $\mathcal{A}$\\
		\textsc{Output:} Cluster head mapping vector $\mathcal{M}$ that denotes the index of the device that acts as cluster head of each device}
		\State $\mathcal{M} \gets \emptyset$
		\For{each device $d_i \in \mathcal{D}$}
		\State $\mathcal{M}_i \gets -1$
		\EndFor
		\For{each centroid $c_k \in \mathcal{C}$}
		\State $d_j \gets \textsc{Best\_Device}(\mathcal{D}, \mathcal{M}, c_k)$
		\State \textsc{Associate\_Devices}($\mathcal{A}, \mathcal{M}, d_j$)
		\EndFor
		\EndProcedure \\
		
		\Procedure{Best\_Device}{$\mathcal{D}$, $\mathcal{M}$, $c$}\\
		{\color{darkgray}\textsc{Input:} Set of devices $\mathcal{D}$, cluster head mapping vector $\mathcal{M}$, and specific centroid $c$ \\
		\textsc{Output:} Best device that can play the role of the input virtual centroid $c$} 
		\State best\_device $\gets \emptyset$
		\State min\_distance $\gets \infty$
		\For{each device $d_i \in \mathcal{D}$}
		\State distance $\gets$ \textsc{Distance}($c.x$, $c.y$, $d_i.x$, $d_i.y$)
		\If{$\mathcal{M}_i$ == $-1$ AND distance $<$ min\_distance}
		\State min\_distance $\gets$ distance
		\State best\_device $\gets$ $d_i$
		\EndIf
		\EndFor
		\State \Return best\_device
		\EndProcedure \\
		
		\Procedure{Associate\_Devices}{$\mathcal{A}, \mathcal{M}, d_j$}\\
		{\color{darkgray}\textsc{Input:} Association matrix $\mathcal{A}$, cluster head mapping vector $\mathcal{M}$, and device $d_j$ \\
		\textsc{Output:} Updated cluster head mapping vector $\mathcal{M}$ to indicate $d_j$ as the cluster head of all devices assigned to the corresponding virtual centroid, where $d_j$ is output by  \textsc{Best\_Device}}
		\State $\mathcal{M}_j \gets j$
		\For{each device $d_i \in \mathcal{A}_j$}
		\State $\mathcal{M}_i \gets j$
		\EndFor
		\EndProcedure\\
	\end{algorithmic}
\end{algorithm}

\begin{algorithm}[th!]
	\caption{\textsc{RForce - Phase 3}}
	\label{RForce-MST}
	\begin{algorithmic}[1]	
		\Procedure{RForce - MST}{$\mathcal{S}$, $\mathcal{H}$}\\
		{\color{darkgray}\textsc{Input:} Set of access points $\mathcal{S}$, and set of cluster heads $\mathcal{H}$\\
		\textsc{Output:} Association matrix $\mathcal{B}$ that maps each cluster head to one access point}
		\State $Q \gets \emptyset$
		\State $\mathcal{B} \gets \emptyset$
		\For{each $s \in \mathcal{S}$}
		\For{each $h \in \mathcal{H}$}
		\State $Q$.push($s, h$)
		\EndFor
		\EndFor
		\While{$Q$ is not empty}
		\State	$s, h$ = $Q$.pop()
		\If{$s$.degree $< \Delta_{\mathrm{LR}}$}
		\State $\mathcal{B}_h \gets s$
		\State $s$.degree $\gets s$.degree $+ 1$
		\EndIf	
		\EndWhile
		\State \Return $\mathcal{B}$
		\EndProcedure \\
	\end{algorithmic}
\end{algorithm}

Algorithm~\ref{RForce-P1} is called with two main inputs, a vector of randomly generated centroids and a vector of devices with given locations; the devices are the set of sensors or users deployed in the area of interest. The algorithm then goes into a loop that only halts when the centroids reach a stable state whereby their positions are nearly no longer varying. Inside the loop, the algorithm initially starts by resetting all associations between centroids and devices that have been done in the previous iteration by setting the centroids degree to zero and finding the best set of devices for each centroid. To do this, the algorithm utilizes \textsc{Associate\_Centroids} method which loops through the devices and finds for each device the nearest centroid using the \textsc{Best\_Centroid} method taking into account the constraint on degree bound. After finding the best centroid for each device, each centroid $c_k$ gets its own vector of associated devices which form the set ${\cal A}_k$; hence, ${\cal A}_k$ with $c_k$ form an initial cluster. After obtaining the corresponding centroid of each device, we can now calculate the forces exerted on each centroid by the set of devices within its cluster and by the other centroids, using ~(\ref{eq:sumForces}). Then the centroid is moved according to~(\ref{eq:force_step}).

In Phase~2 of the algorithm, the virtual centroids with optimized locations are mapped to existing devices that will act as cluster heads using the intelligence summarized in Algorithm~\ref{RForce-P2}. It starts by defining a set $\mathcal{M}$ that stores the indices of the cluster head node for each device $d_i$. Initially, $d_i$ is set to -1 which means that this set does not belong to any group. We then start by looping on the set of centroids and finding for each virtual centroid a corresponding device that is not connected to a cluster and is within minimal distance to achieve high communications quality, using the \textsc{Best\_Device} method. Finally, after mapping each virtual centroid to a cluster head device, we shift all the virtual centroid's associated devices to the corresponding cluster head device to form a cluster.


In Phase~3 of the algorithm, all cluster heads need to be connected to the available set of APs as our system model assumes an infrastructure based network scenario. The allocation should maximize performance quality while respecting the constraint on the number of LR connections  per AP. One can address this using a standard MST~(Minimum Spanning Tree) algorithm such as Prim or Kruskal in order to ensure that one cluster head receives its content from only one of the available APs. However, since the APs have a degree bound of $\Delta_{\mathrm{LR}}$, this problem reduces to MBDST~(Minimum Bounded Degree Spanning Tree) which is known to be NP-hard \cite{4031363}. Thus, we resort to a modified version of Kruskal's algorithm to associate cluster heads to APs without violating the degree constraint of the APs and the tree property (see Algorithm~\ref{RForce-MST} for the details).

\section{Performance Results and Analysis}
\label{sec:results}

In this section, we demonstrate the effectiveness of the proposed proactive and scalable approach for reliable cluster formation in cooperative wireless networks using a combination of optimization problem solutions for relatively small scale network scenarios, Monte-Carlo simulation results based on the presented low complexity algorithms in Section~\ref{sec:method}, and experimental test bed results to demonstrate feasibility under realistic operational conditions. For the optimization problem solution, we use the intlinprog mixed integer linear programming function in Matlab (denoted as $\mathrm{ROptimal}$ in the sequel). We also compare the $\mathrm{RForce}$ algorithm to the standard $\mathrm{kMeans}$ clustering algorithm in terms of reliability, download bit rates, and execution complexity. For $\mathrm{kMeans}$ clustering, we use Lloyd's algorithm~\cite{lloyd} with slight modification to capture our problem constraints. We consider several network scenarios and vary system parameters to produce a wide range of results that allow for extracting insights and capturing tradeoffs. For the Monte-Carlo simulation results, we average over 25 runs for each set of network scenario and system parameters and plot average performance metrics. 
Table~\ref{tab:constants_used} summarizes the values used in the results for key network and algorithm parameters.

\begin{table}[!htb]
	\caption{Network and algorithm parameters}
	\centering
	\begin{tabular}{ |l|l| }
		\hline
		Parameter & Value \\
		\hline
        Area size   & 100m by 100m          \\\hline
		AP transmit power      & 10 Watts                 \\\hline
		Device transmit power      & 0.22 Watts               \\\hline
		$\beta$	   & 0.3					  \\\hline
		$\sigma^2$ & $1 \times 10^{-9}$ Watts \\\hline
        $\rho$   & 20                       \\\hline
		$\theta$   & 0.05                     \\\hline
		$\Delta_{\mathrm{LR}}$ & 30 connections per AP        \\\hline
		$\Delta_{\mathrm{SR}}$ & 10 connections per cluster head  \\\hline
		$\mathrm{SNR_{LR}}$  & 1                        \\\hline
        $\mathrm{SNR_{SR}}$  & 1                        \\\hline
		$\lambda$  & 0.8                      \\\hline
		$\eta$     & 0.4                      \\\hline
        $\gamma_i$ $\forall i$  &   1                   \\
		\hline
	\end{tabular}
	\label{tab:constants_used}
\end{table}

\begin{figure}[]
	\centering
	\includegraphics[width=\columnwidth]{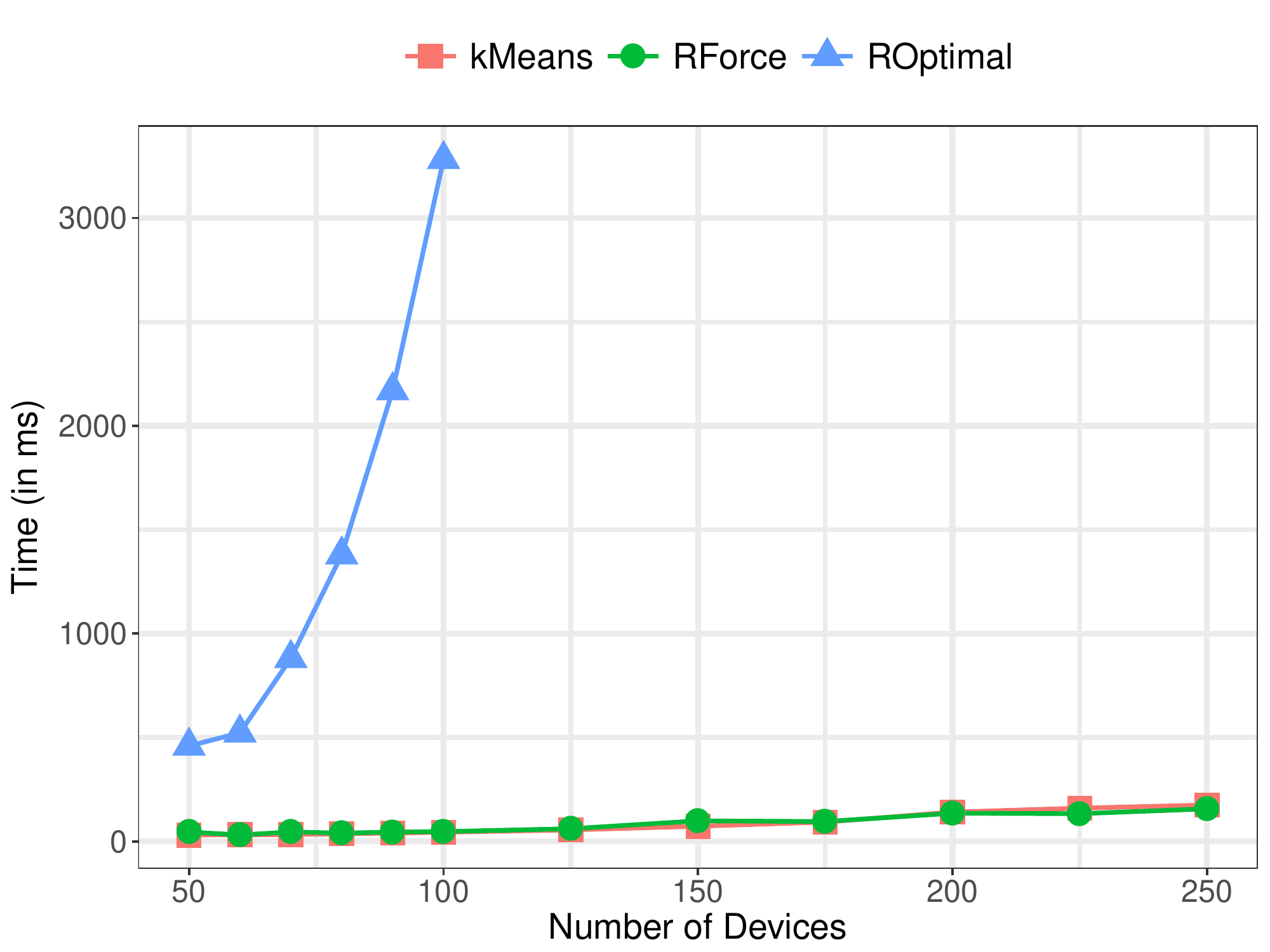}
	\caption{Average execution time of various approaches assuming a network scenario with one AP. }
	\label{fig:uniform_time}
\end{figure}

Fig.~\ref{fig:uniform_time} compares the running time in msec on a standard PC for the different approaches assuming a network scenario with one AP and number of devices ranging from 50 to 250. This clearly demonstrates the gap in complexity between generating the optimal solution and generating solutions using our proposed $\mathrm{RForce}$ algorithm and the standard $\mathrm{kMeans}$ algorithm; the computational complexity of the optimal solution increases exponentially with feasible outcome only for network scenarios having up to 100 devices. On the other hand, $\mathrm{RForce}$ and $\mathrm{kMeans}$ are shown to have similar low complexity and fast execution time which makes both of them applicable to ultra dense network scenarios with dynamic adaptation over time.

\begin{figure}[h!]
 \begin{subfigure}{0.5\textwidth}
\includegraphics[width=1\linewidth, height=7cm]{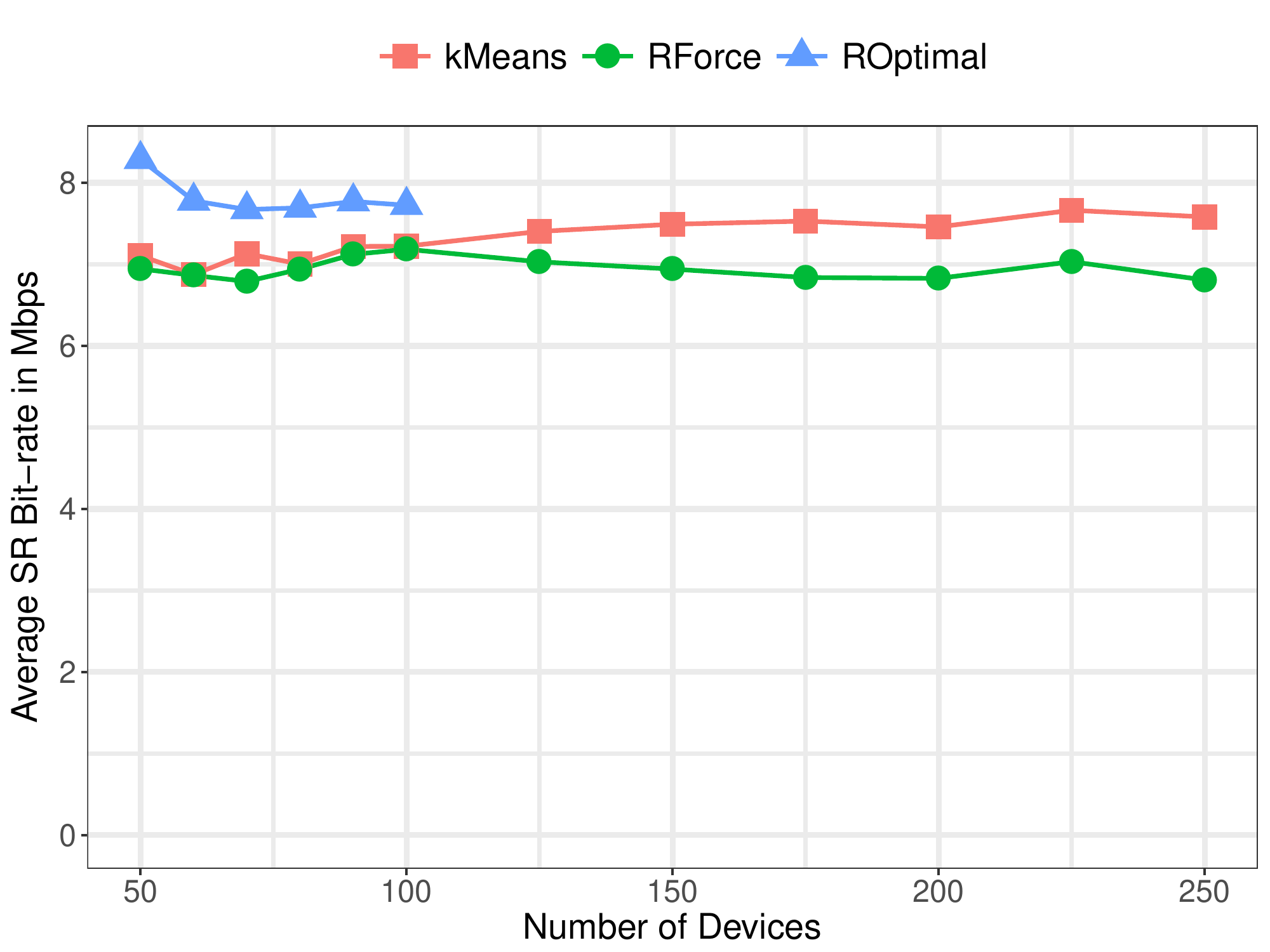}
\end{subfigure}
\begin{subfigure}{0.5\textwidth}
\includegraphics[width=1\linewidth, height=7cm]{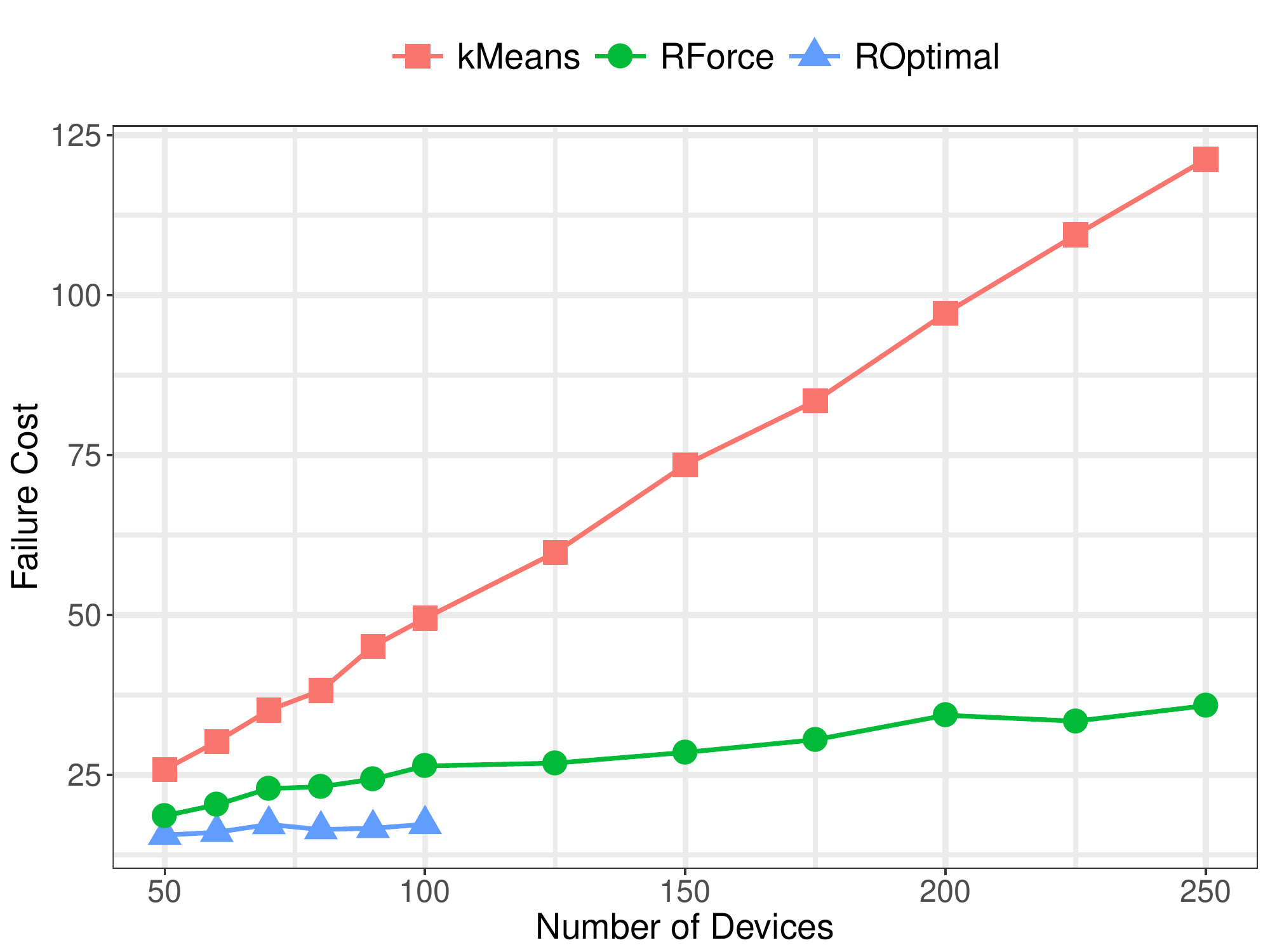}
\end{subfigure}
 \caption{Top: Average network SR bit rate in Mbps versus number of devices assuming one AP; Bottom: Average failure cost versus number of devices assuming one AP.}
\label{fig:oneBS}
\end{figure}

\begin{figure}[h!]
 \begin{subfigure}{0.5\textwidth}
\includegraphics[width=1\linewidth, height=7cm]{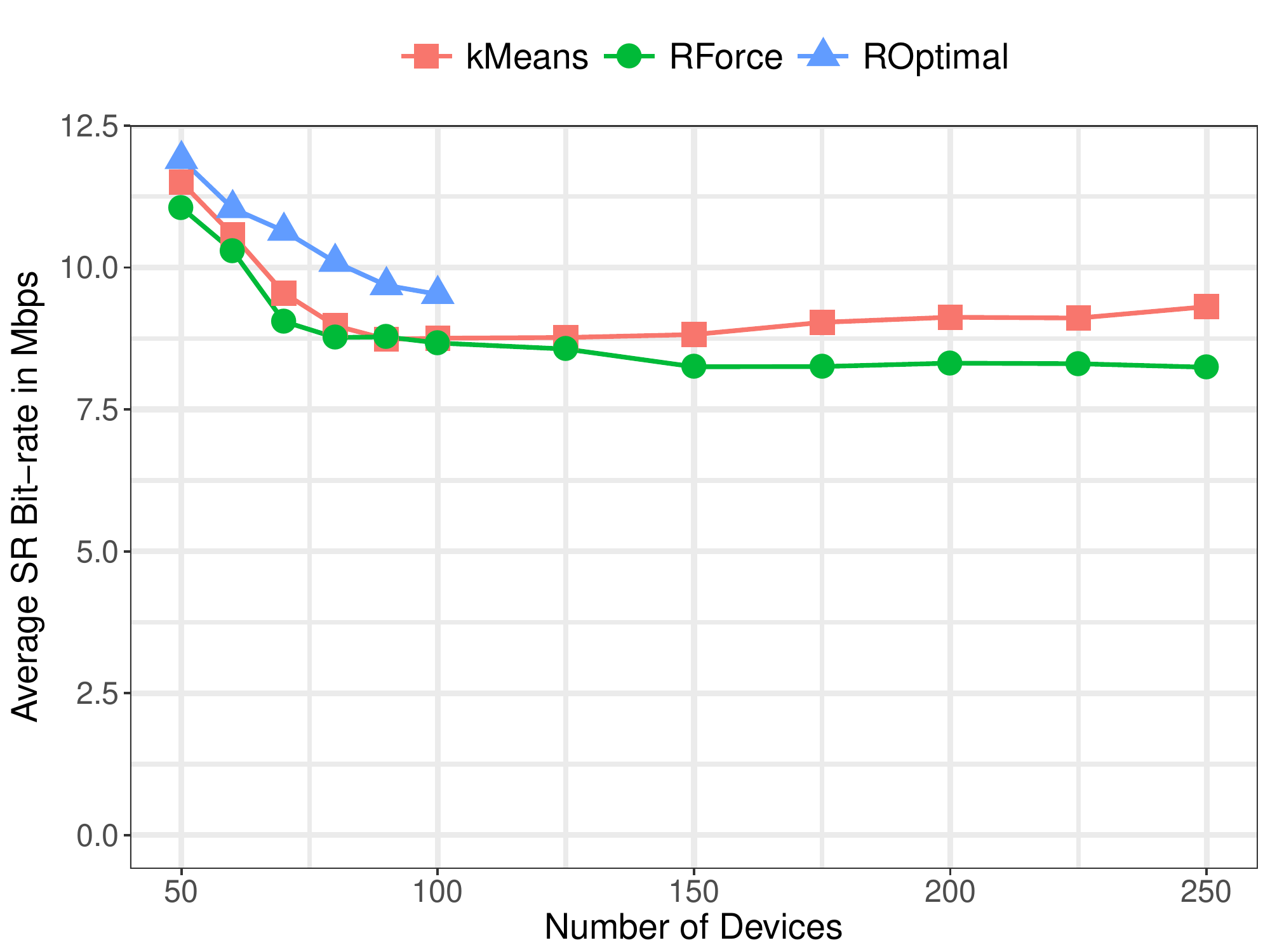}
\end{subfigure}
\begin{subfigure}{0.5\textwidth}
\includegraphics[width=1\linewidth, height=7cm]{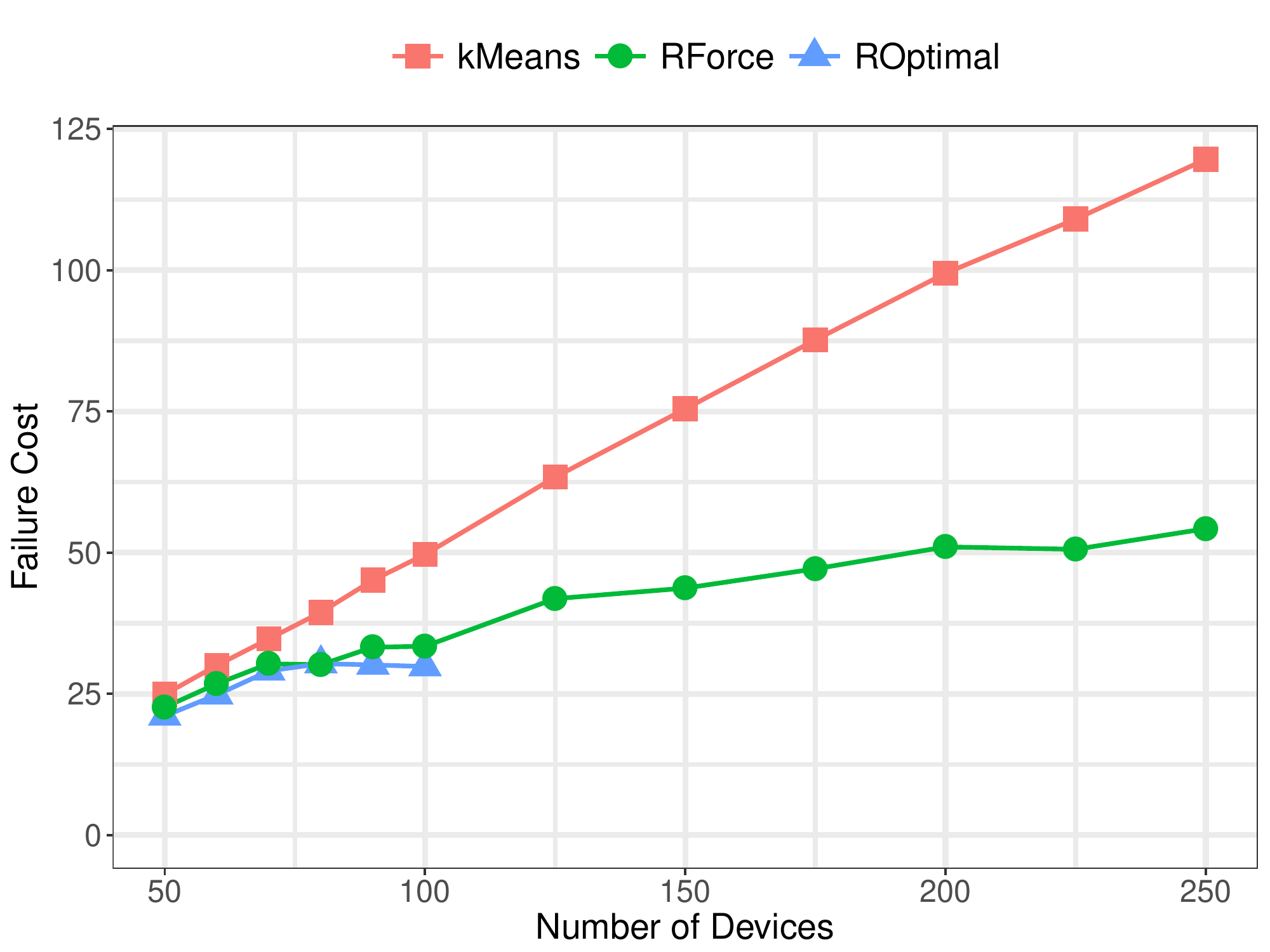}
\end{subfigure}
 \caption{Top: Average network SR bit rate in Mbps versus number of devices assuming four APs; Bottom: Average failure cost versus number of devices assuming four APs.}
\label{fig:fourBS}
\end{figure}

Fig.~\ref{fig:oneBS} and Fig.~\ref{fig:fourBS} compare the performance of the different algorithms in terms of communications quality and reliability for network scenarios with one AP and four APs, respectively, assuming the number of devices to be served is between 50 and 250. The results demonstrate the effectiveness of the proposed $\mathrm{RForce}$ approach in achieving network reliability close to the optimal solution and much higher than the $\mathrm{kMeans}$ approach; this is reflected in the plots showing the failure cost metric, defined in~(\ref{cost}), as a function of the number of devices. The results also demonstrate that this effectiveness in reliability is not negatively impacting communications bit rate or download speed, as the average bit rate on the SR links is shown to be close between $\mathrm{RForce}$ and $\mathrm{kMeans}$, with both worse than the optimal reliable clustering solution. For example, assuming 200 devices in a given area with one AP, clusters formed by $\mathrm{kMeans}$ lead to around triple the failure cost compared to the solution produced by $\mathrm{RForce}$, at the expense of around 0.5~Mbps reduction in average SR bit rate (around 10\% reduction only); this is expected since $\mathrm{kMeans}$ aims at maximizing network throughput without accounting for reliability. When compared to the optimal solution, $\mathrm{RForce}$ is shown to have relatively close performance for both bit rate and reliability, yet with much higher execution speed as shown in Figure~\ref{fig:uniform_time}.

The same trends and insights on the effectiveness of  $\mathrm{RForce}$  are also demonstrated in Fig.~\ref{fig:fourBS} with four APs; compared to the results with one AP, the bit rates become higher since more devices are served by the four APs on LR connections with closer proximity, whereas the failure cost also increases since more devices are chosen as cluster heads due to their proximity to the APs and some of these chosen cluster heads do not have high reliability metrics. These observations are consistent with the tradeoff between bit rate and reliability as captured in the objective function of the optimization problem formulation in~(\ref{obj_function}).

\begin{figure}[h]
 \begin{subfigure}{0.5\textwidth}
\includegraphics[width=1\linewidth, height=7cm]{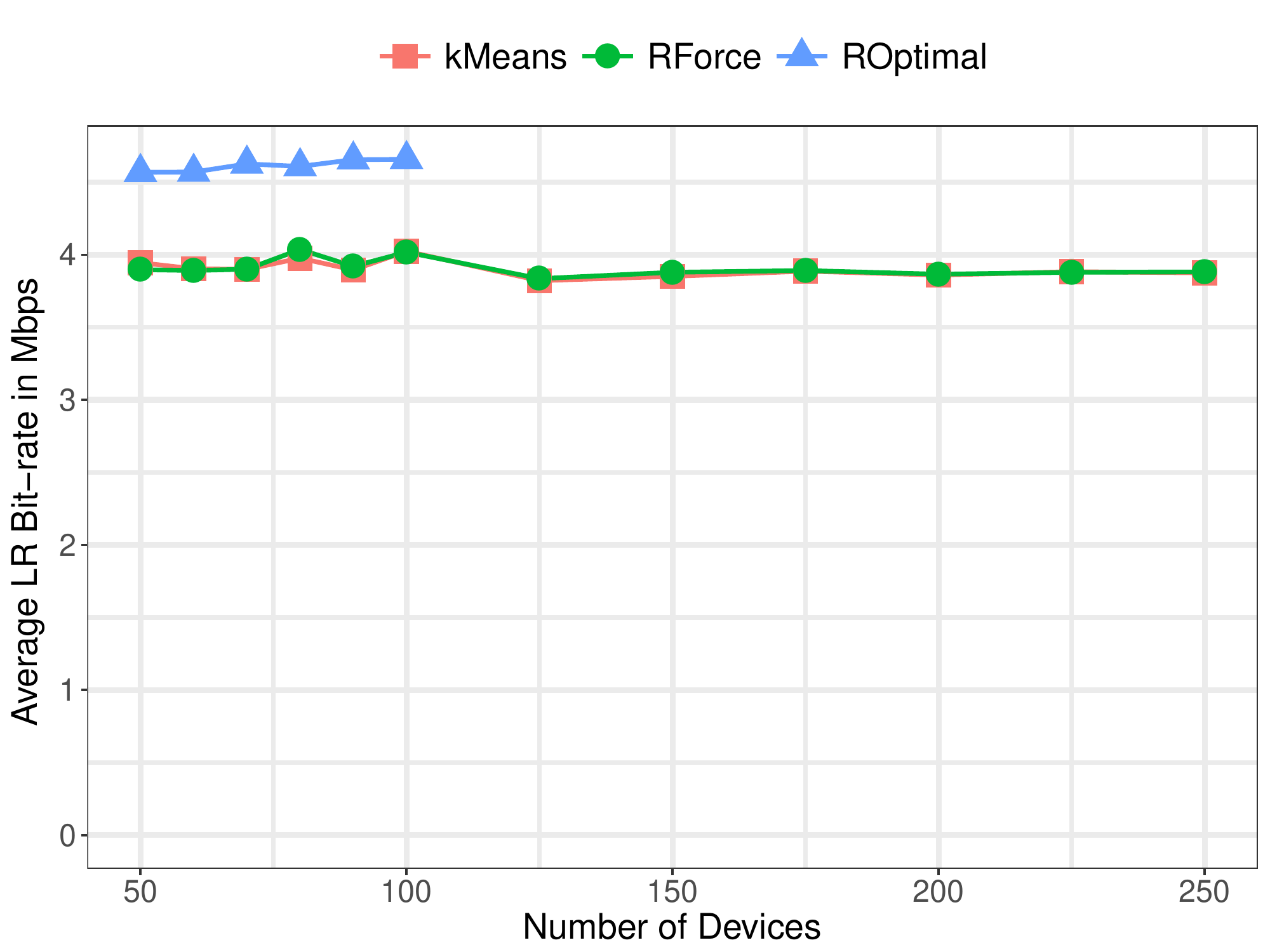}
\end{subfigure}
\begin{subfigure}{0.5\textwidth}
\includegraphics[width=1\linewidth, height=7cm]{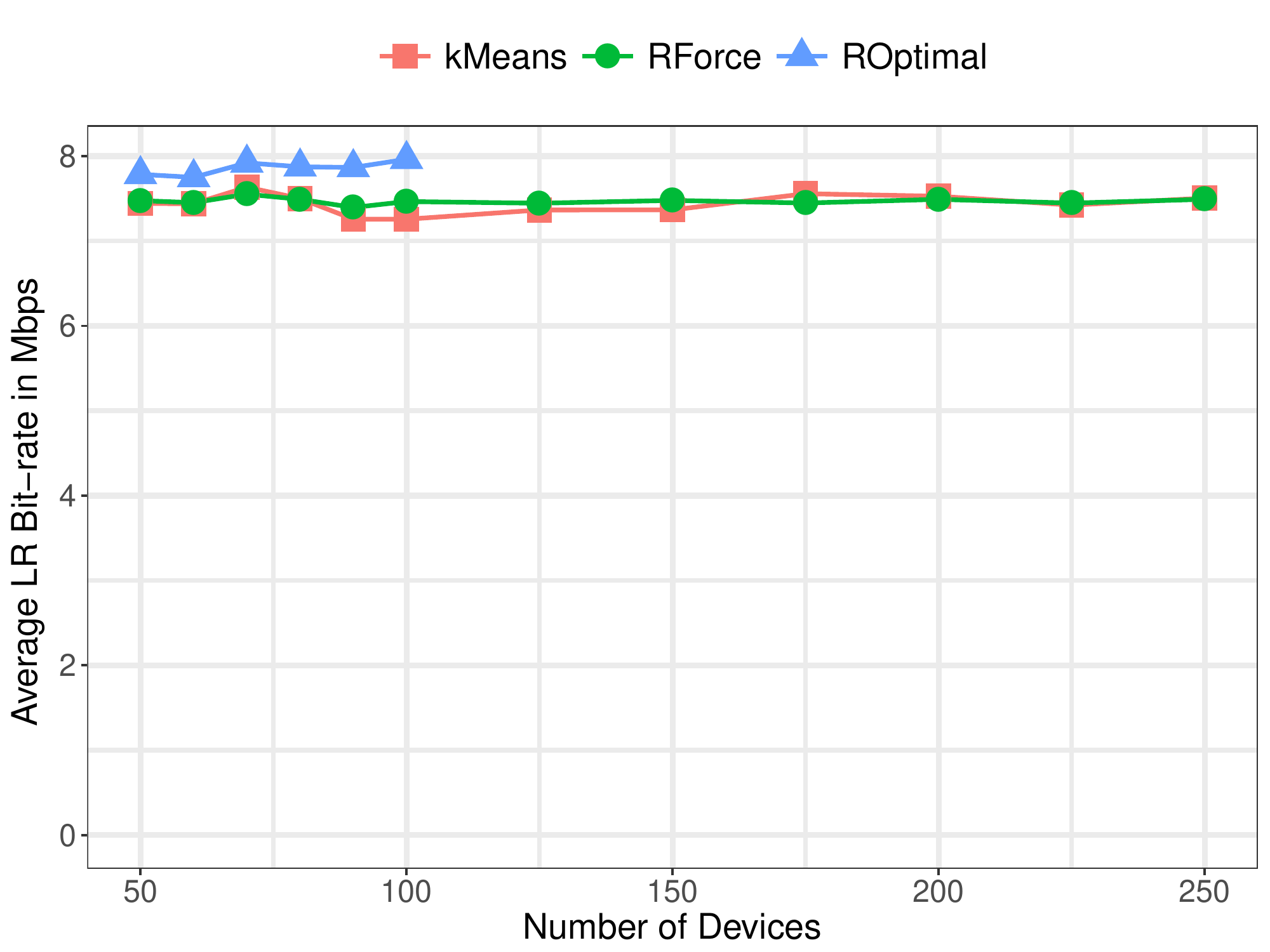}
\end{subfigure}
\caption{Top: Average network LR bit rate in Mbps versus number of devices assuming one AP; Bottom: Average network LR bit rate in Mbps versus number of devices assuming four APs.}
\label{fig:lr-bitrates}
\end{figure}

We have also compared the various algorithms in terms of average network LR bit rate with results shown in Fig.~\ref{fig:lr-bitrates} assuming network scenarios with one and four APs having between 50 and 250 devices. Similar to the SR bit rates, the results produced by $\mathrm{RForce}$ are almost equivalent to $\mathrm{kMeans}$ and both are not far from $\mathrm{ROptimal}$. These results demonstrate the effectiveness of Algorithm~\ref{RForce-MST} in associating cluster heads with APs with high bit rates on the LR connections.

\begin{figure}[t!]
 \begin{subfigure}{0.5\textwidth}
\includegraphics[width=1\linewidth, height=7cm]{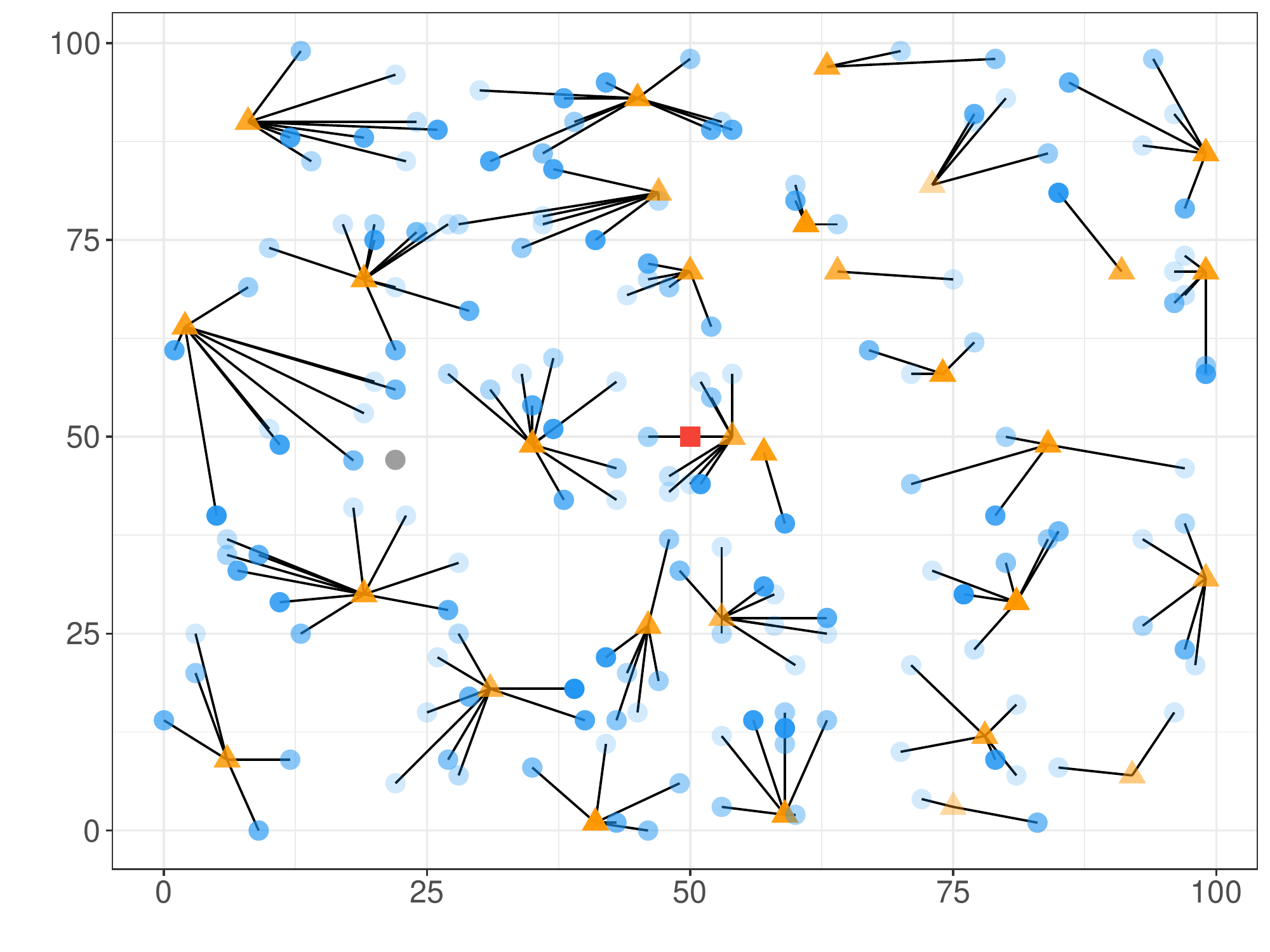}
\end{subfigure}
\begin{subfigure}{0.5\textwidth}
\includegraphics[width=1\linewidth, height=7cm]{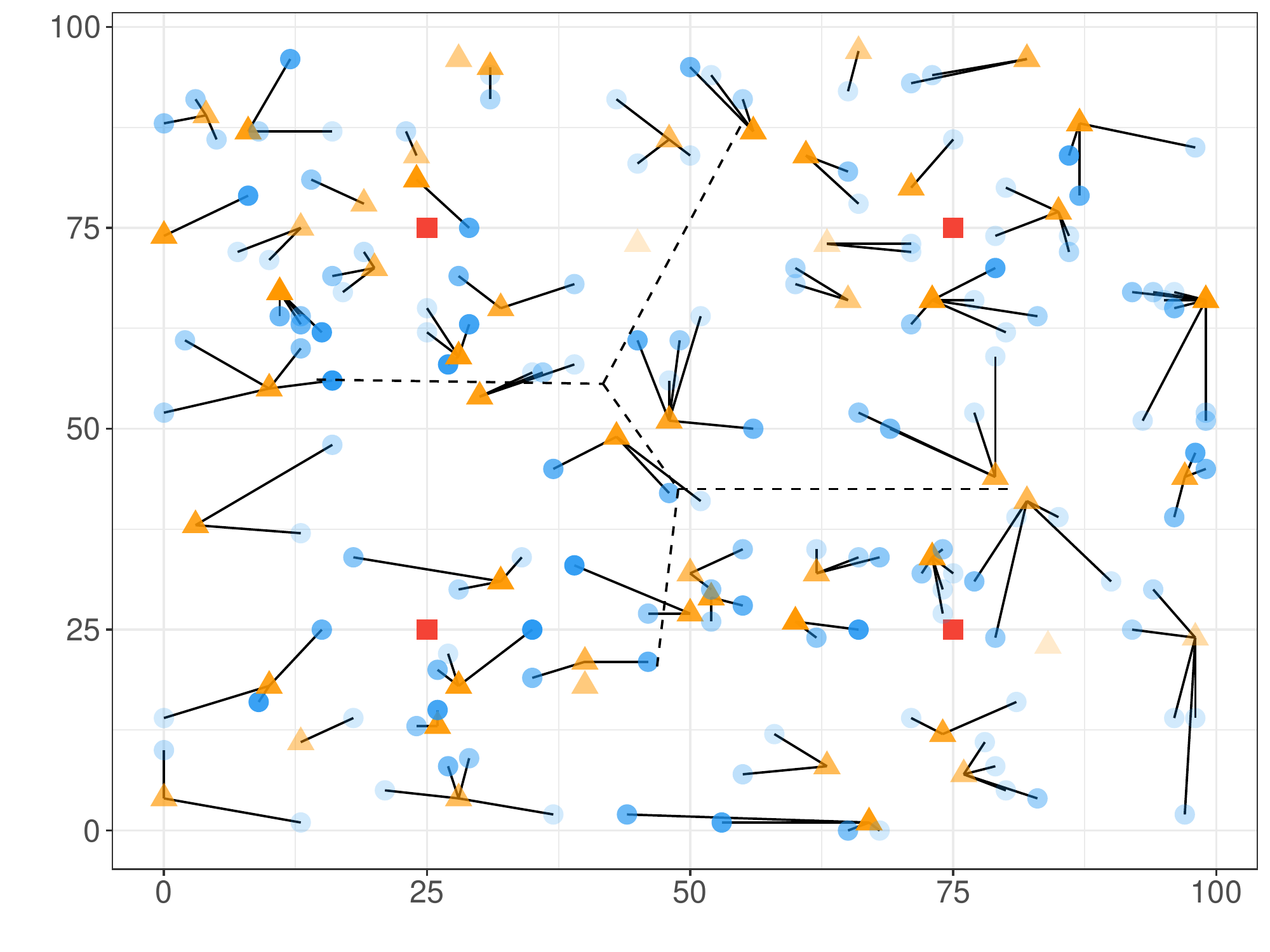}
\end{subfigure}
 \caption{Sample snapshot network scenarios generated by $\mathrm{RForce}$ assuming one AP (upper plot) and four APs (lower plot) with 200 devices. }
\label{fig:scenarios} 
\end{figure}

\begin{figure}[t!]
 \begin{subfigure}{0.5\textwidth}
\includegraphics[width=1\linewidth, height=7cm]{area_RForce.pdf}
\end{subfigure}
\begin{subfigure}{0.5\textwidth}
\includegraphics[width=1\linewidth, height=7cm]{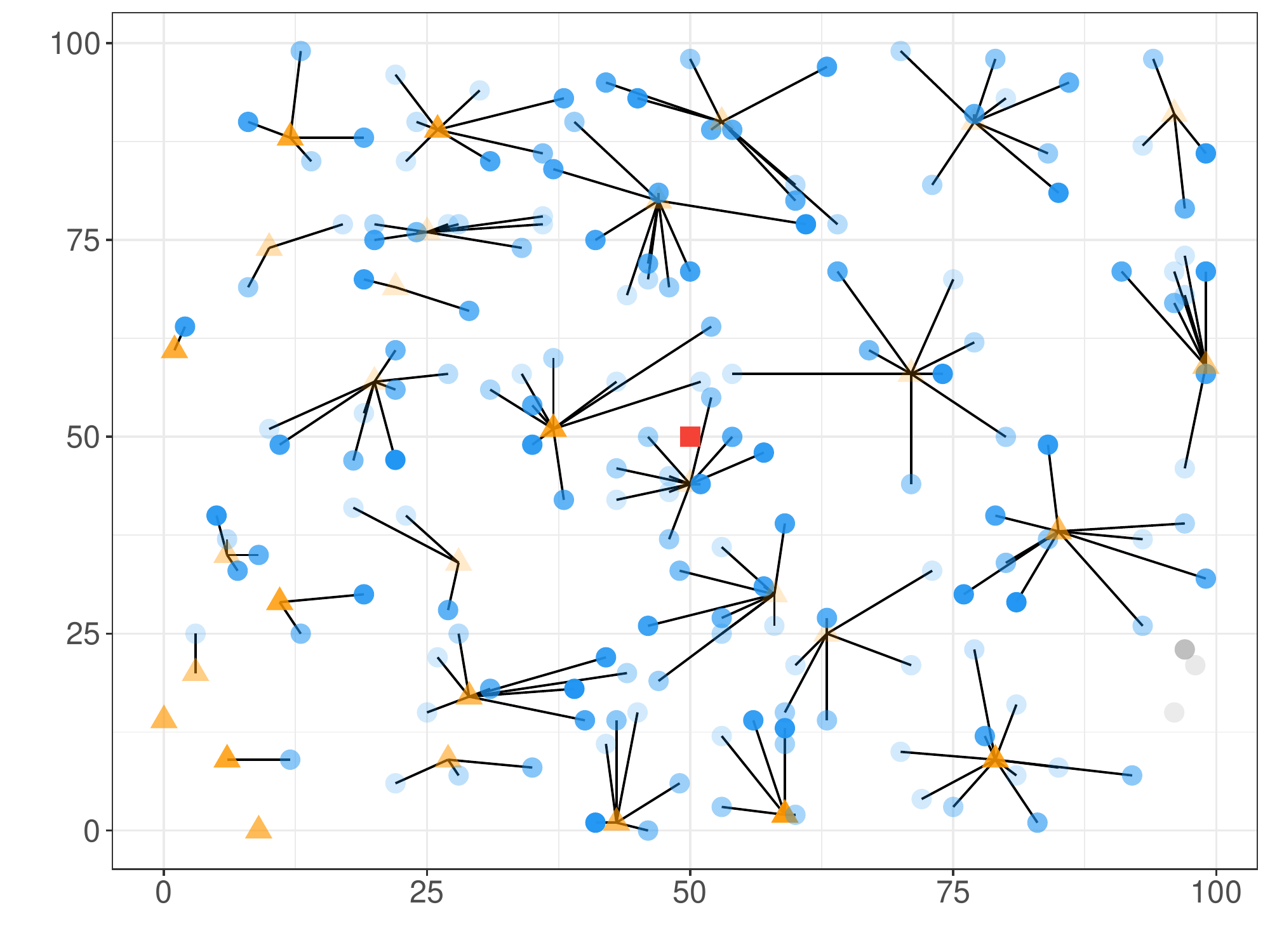}
\end{subfigure}
\caption{Sample snapshot network scenarios assuming one AP and 200 devices with clustering using $\mathrm{RForce}$ (upper plot) and $\mathrm{kMeans}$ (lower plot). }
\label{fig:scenarios2}
\end{figure}

Next, we present sample snapshot network scenarios with results generated using $\mathrm{RForce}$ to provide additional insights on number of clusters, their locations, and selection of cluster heads. Fig.~\ref{fig:scenarios}a presents one sample netwrok scenario with 200 devices uniformly distributed in a $100m\times100m$ area. We represent APs as red squares, cluster heads as orange triangles, connected devices as blue circles, and unconnected devices in outage as gray circles. We also designate the reliability factor of each device by its opacity level. The figure shows that highly reliable devices (darker colors) are selected as cluster heads and they are also well positioned among the devices they serve in their cluster. Fig.~\ref{fig:scenarios}b shows another example scenario with four APs and 200 devices. We notice that $\mathrm{RForce}$ tends to select more cluster heads when the number of APs increases as the algorithm tends to choose more devices as cluster heads to be served directly by the APs due to their proximity. It is shown also that there are clusters composed of a single device served by in its own by the AP in order to avoid an increase in outage rate. Fig.~\ref{fig:scenarios2} compares the resulting clusters between $\mathrm{RForce}$ and $\mathrm{kMeans}$ approaches; these plots show clearly the relative limitations of $\mathrm{kMeans}$ as devices with low reliability are chosen as cluster heads and outage rate is higher with more devices not served.
	
\subsection{Energy Efficiency and Scalability}

In order to provide a more tangible quantification of the impact of the failure cost metric, we estimate the lifetime of each selected cluster head based on a given set of assumptions taking into account available battery budget. We consider devices powered with batteries of $2000~mAh$ capacity and use the battery consumption model presented in~\cite{Halperin:2010:DPC:1924920.1924928}; to this end, we assume that devices are downloading data continuously with power consumption $1.27~W$. We then derive the current in Ampere to be $0.34~A$ assuming a nominal voltage of $3.7$~V. Battery lifetime is then calculated as $E*2000~mAh/0.34~A$, where $E$ is the actual battery indicator in percentage. In our simulations, we assigned to the devices arbitrary values of $E$ ranging from 10\% to 90\% to mimic a realistic network scenario.

\begin{figure}[htb!]
	\centering
	\includegraphics[width=\columnwidth]{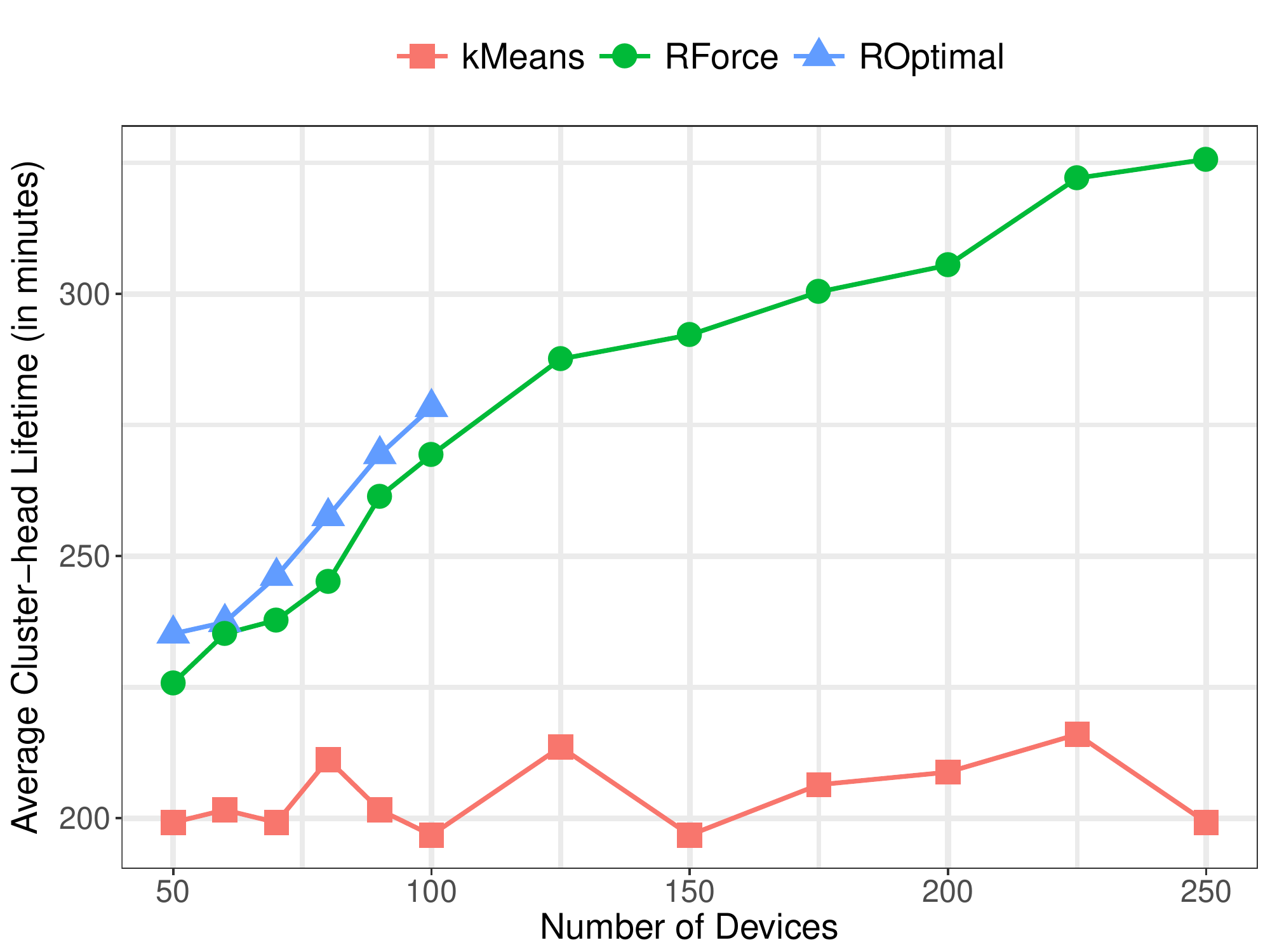}
	\caption{Average cluster head lifetime in minutes as a function of the number of devices assuming a network scenario with one AP. }
	\label{fig:uniform_lifetime}
\end{figure}

Fig.~\ref{fig:uniform_lifetime} plots the resulting average lifetime of all cluster heads selected by each algorithm. $\mathrm{RForce}$ outperforms $\mathrm{kMeans}$ as the number of devices varies with close performance to $\mathrm{ROptimal}$. We note that the average cluster head lifetime of both $\mathrm{RForce}$ and $\mathrm{ROptimal}$ increase as the number of devices increases within the given area due to the fact that they tend to select the most reliable cluster heads and more of these become available as the number of devices increases in the given area of interest. The performance of $\mathrm{kMeans}$, however, is almost constant with limited variability as the number of devices increases since the algorithm does not account for reliability during the clustering process.

\begin{figure}[htb!]
	\centering
	\includegraphics[width=\columnwidth]{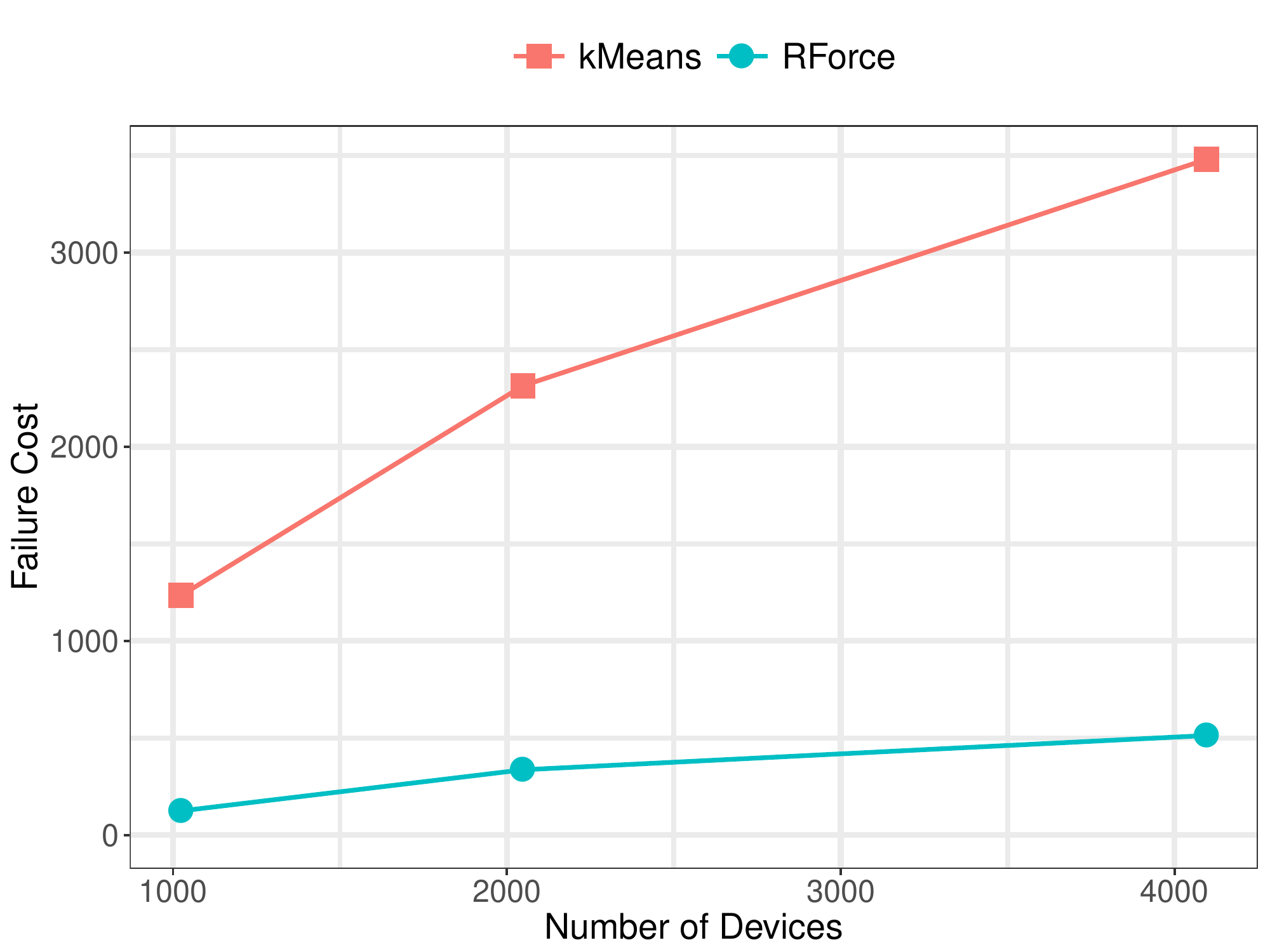}
	\caption{Average failure cost for ultra dense network scenarios with up to 4000 devices. }
	\label{fig:scalability} 
\end{figure}

Finally, Fig.~\ref{fig:scalability} presents results for ultra dense network scenario with up to 4000 devices in order to demonstrate the scalability of the proposed $\mathrm{RForce}$ approach and reconfirm its superiority in terms of reliability compared to the standard $\mathrm{kMeans}$ approach. It is important to note that it was not feasible to generate optimal results for such high density network scenarios due to high complexity.

\subsection{Experimental Evaluation using Test Bed Measurements}

In this section, we demonstrate the feasibility of implementing the $\mathrm{RForce}$ algorithm in a real experimental test bed and evaluate its operation under realistic network conditions. Our test bed consists of a set of devices that are connected via a WiFi access point to download content from a remote server. The devices run software applications that facilitate formation of clusters with coordination from a centralized management server. In our implementation, we used Android smartphones to mimic the devices as a proof of concept since smartphones can be programmed to communicate over two interfaces and serve as relays in cooperative networks, e.g., using WiFi or cellular over LR with APs and WiFi-Direct or Bluetooth over SR directly with neighboring devices. Similar implementation can also be done on other types of devices such as sensors or IoT nodes.

\begin{figure}[htb!]
	\centering
	\includegraphics[width=\columnwidth]{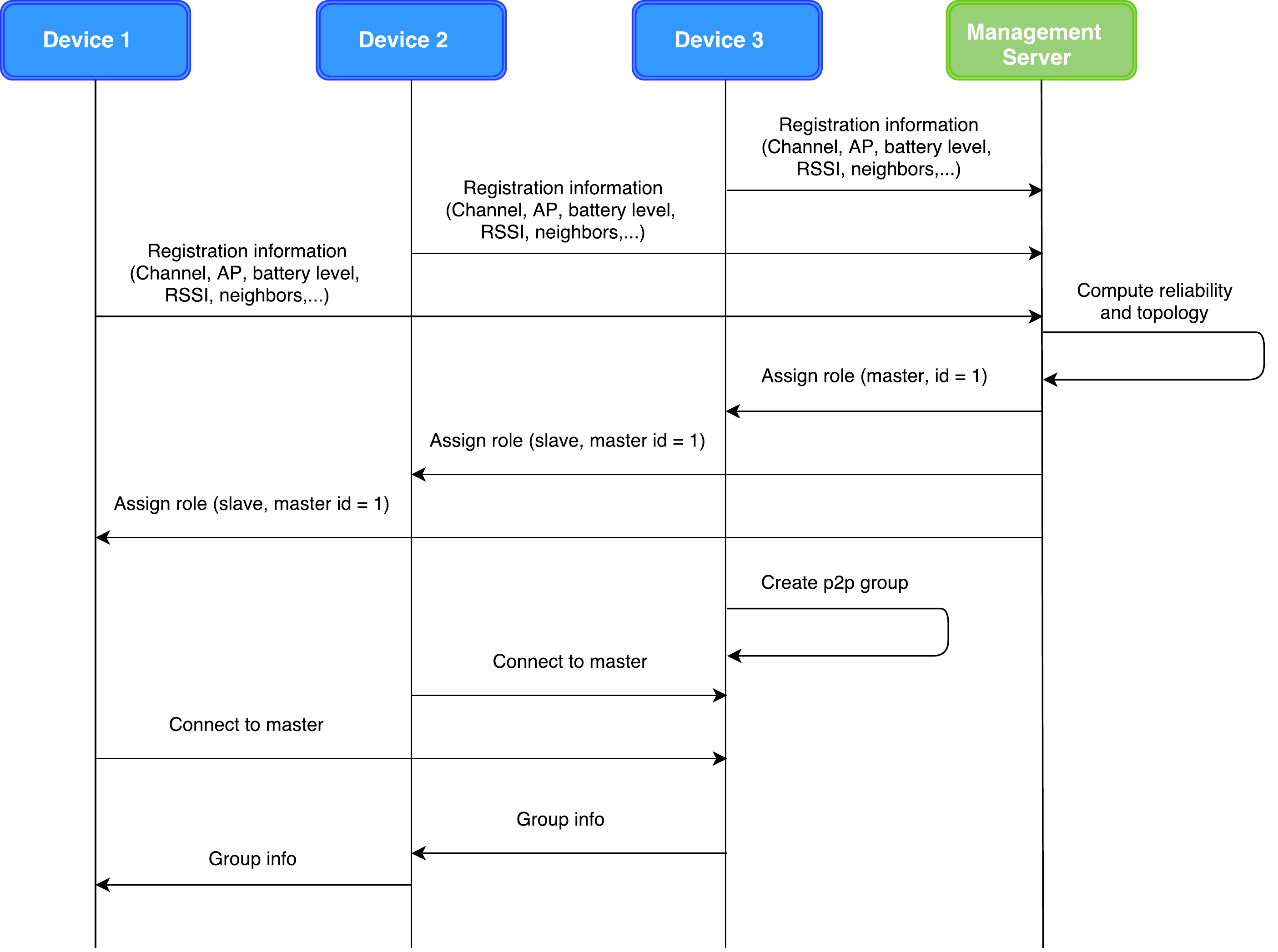}
	\caption{Flowchart showing summary of message exchanges between devices and the management server in our testbed implementation.}
	\label{fig:arch}
\end{figure}

The software running at the device level starts with a discovery phase where all devices scan the area and report to the management server various information including their IDs (e.g., MAC addresses), current battery capacity, list of discovered neighboring devices, and the captured received signal strength level from each neighboring device and AP. Fig.~\ref{fig:arch} presents a summary of the messages exchanged between a set of devices and the management server.

\begin{figure}[htb!]
	\centering
	\includegraphics[width=10cm]{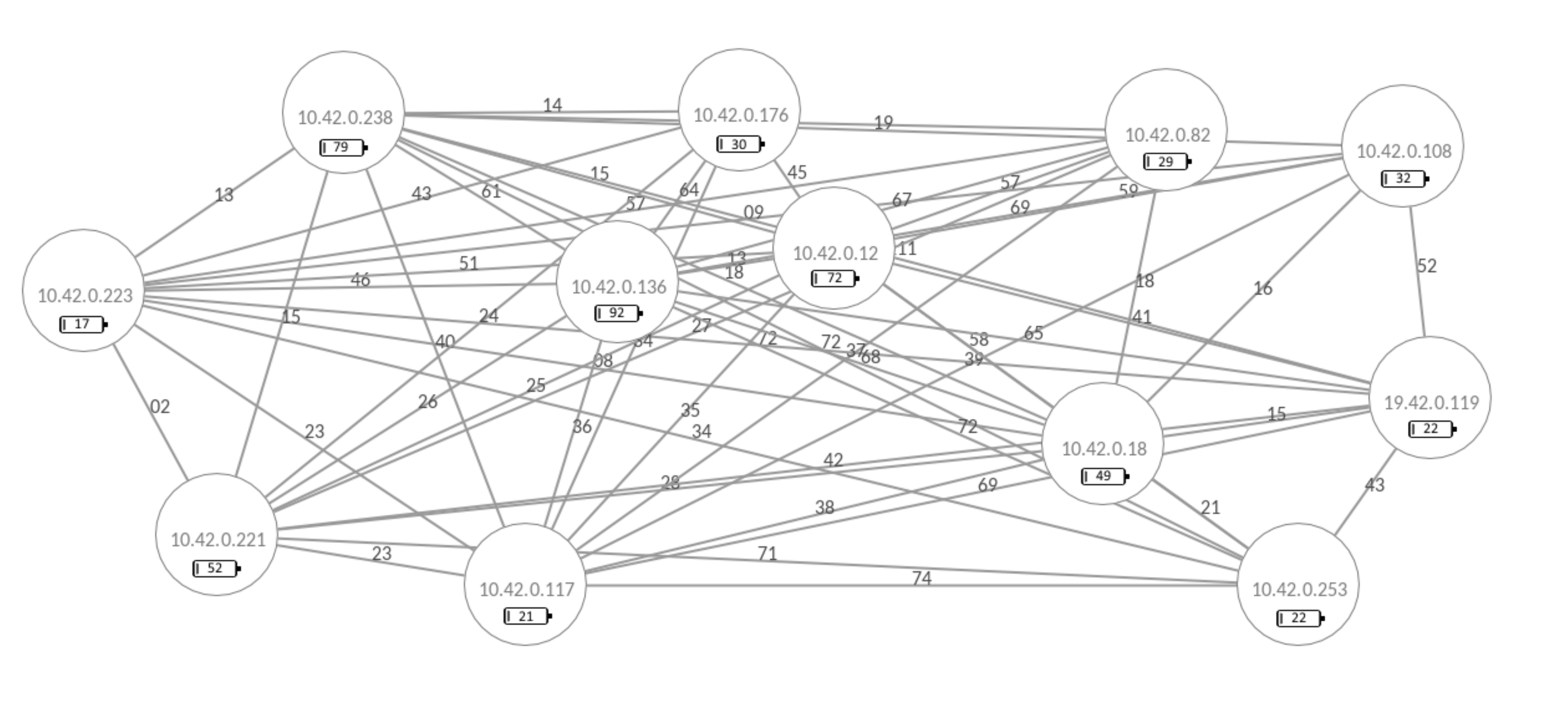} 
	\caption{Snapshot from the test bed server showing the initial graph with a total of 12 devices that need to stream  content from a remote server.}
	\label{fig:implementation_original_1} 
\end{figure}

\begin{figure}[htb!]
	\centering
	\includegraphics[width=10 cm]{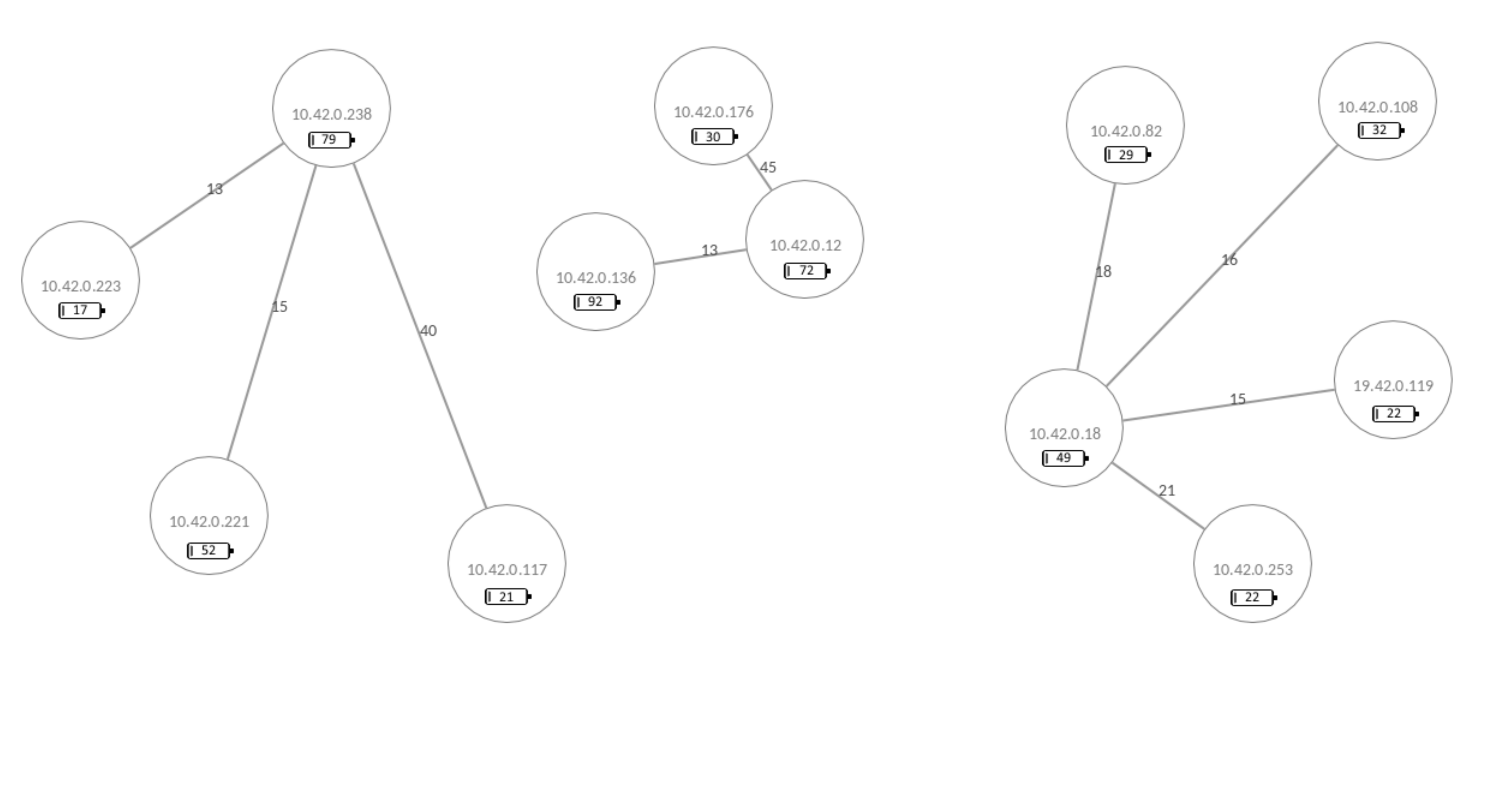} 
	\caption{Snapshot from the test bed server showing the clusters formed after executing the $\mathrm{RForce}$ algorithm.}
	\label{fig:implementation_reduced_1}
\end{figure}

The server uses the received information from the existing devices to build an initial graph of the network as shown in Fig.~\ref{fig:implementation_original_1} from an example real experiment with 12 devices and one AP. This graph is fully meshed because testing is done in a lab where all devices can see each other; it also shows the ID and battery indicator level as reported by each device in addition to a measure of the channel quality between all pairs of devices (the smaller the value on the lines connecting devices, the better the channel quality). After discovering all devices, the server runs the proposed $\mathrm{RForce}$ algorithm on the graph to form reliable clusters of devices. The resulting outcome is shown in Fig.~\ref{fig:implementation_reduced_1} with the devices divided into three clusters with cluster heads having relatively high battery capacity and, thus, high reliability. Note that in the middle cluster, the cluster head is not the best among the three nodes, but it was selected due to its better channel quality with respect to the AP and to the other devices; this demonstrates the tradeoff characteristics of our approach balancing reliability with communications bit rate.

\begin{figure}[h!]
	\centering
	\includegraphics[width=9.9cm]{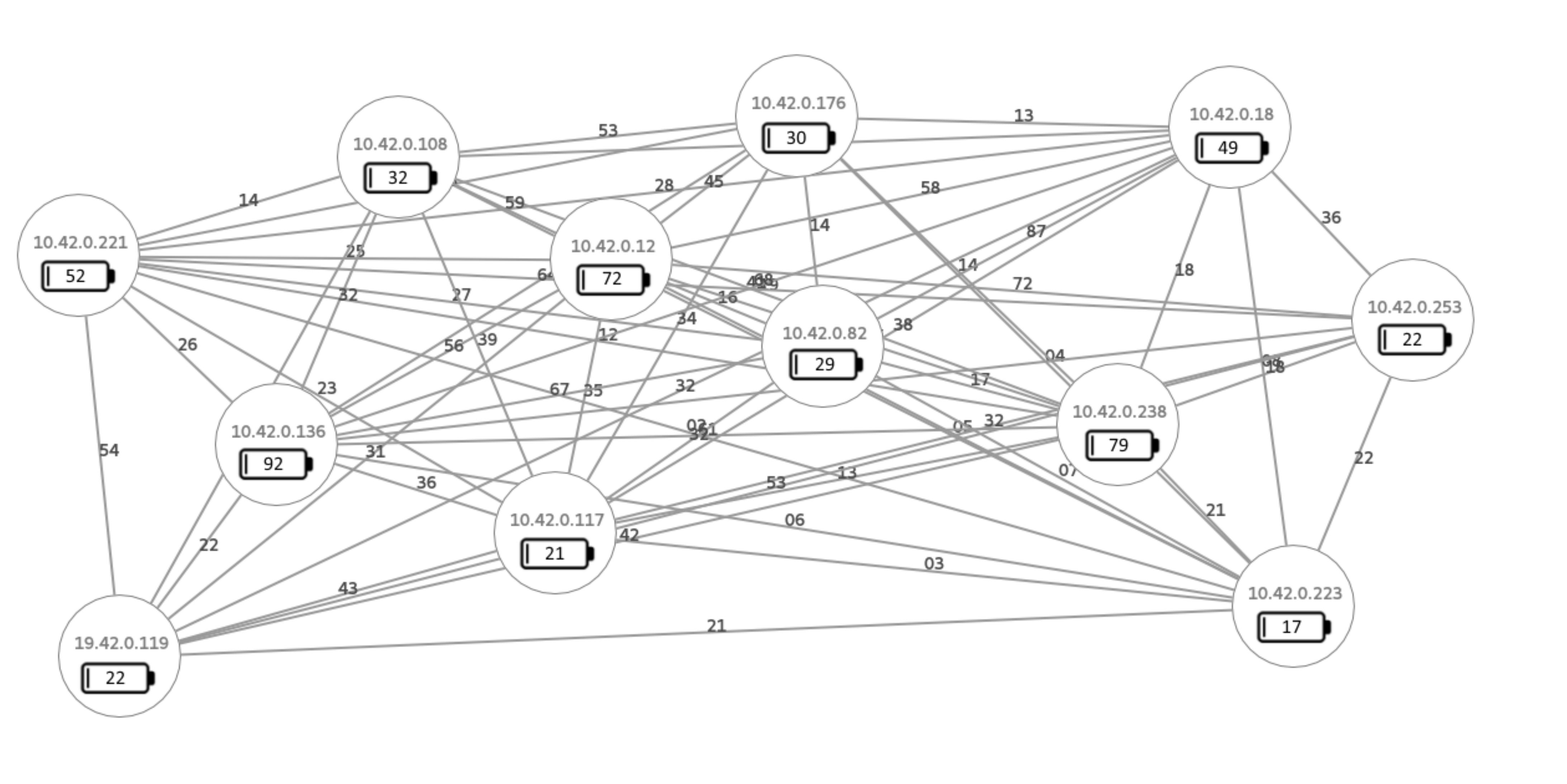}
	\caption{Snapshot from the test bed server showing the initial graph with a total of 12 devices that need to stream  content from a remote server.}
	\label{fig:implementation_original_2}
\end{figure}

\begin{figure}[h!]
	\centering
	\includegraphics[width=9.9 cm]{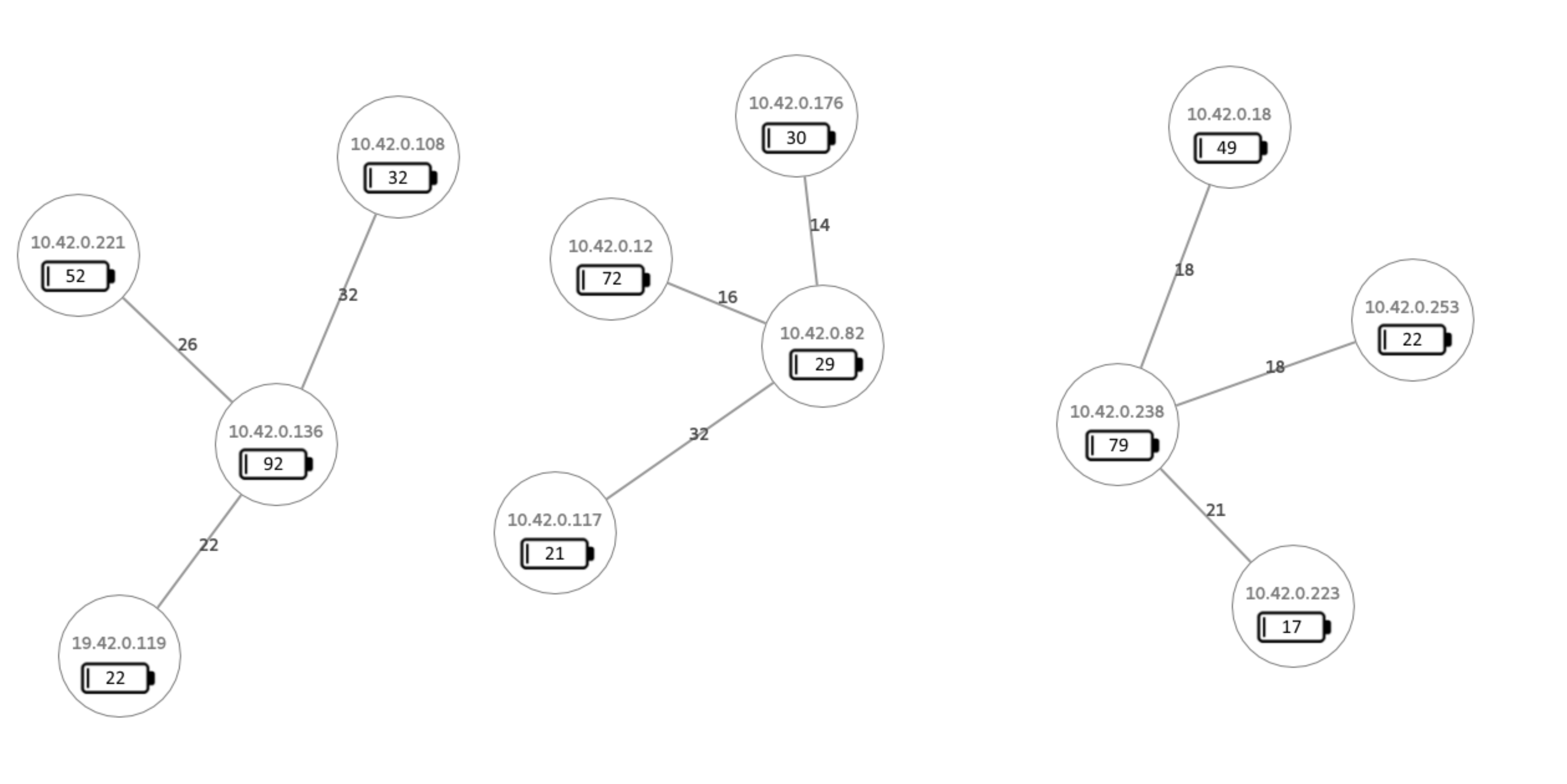}
	\caption{Snapshot from the test bed server showing the clusters formed after executing the $\mathrm{RForce}$ algorithm.}
	\label{fig:implementation_reduced_2}
\end{figure}

  To demonstrate how results vary between different network scenarios, we present in Fig.~\ref{fig:implementation_original_2} and Fig.~\ref{fig:implementation_reduced_2} the initial network graph and the resulting reliable clusters after running the $\mathrm{RForce}$ algorithm, respectively, using another testbed experiment with also 12 devices and one AP.

\section{Conclusions}
\label{sec:conclusions}

In this work, we have proposed a proactive and scalable approach for reliable clustering in wireless networks with device-to-device~(D2D) offloading. Enhancing reliability in cooperative wireless networks is essential for practical implementation as devices (mobile users in a WiFi or cellular network, cars in a vehicular ad hoc network, or sensors in an IoT network) are not controlled by the network operator and, thus, can be dynamic in their mobility, computing and energy resources, and willingness to cooperate. Our proposed approach is characterized as flexible by capturing reliability metrics as part of the clustering process, generic by its application to different wireless network scenarios, and scalable by having low complexity and facilitating dynamic real time implementation. The solution methodology combines integer linear optimization problem formulation with heuristic algorithm design based on the notion of electrostatic forces. We utilize both extensive Monte-Carlo simulations and experimental test bed demonstrations in order to evaluate performance gains with respect to standard techniques and highlight interesting reliability-rate tradeoffs. The results confirm that it is practically feasible to enhance D2D cooperation reliability notably with a limited tradeoff cost in communication bit rate using the proposed efficient reliable clustering approach.


\section*{Acknowledgements}
\label{sec:acknowledgements}
This work is supported by the Lebanese American University under LAU fund \# SRDC-r-2017-4.

\bibliographystyle{ieeetr}
\bibliography{references}

\begin{thebibliography}{10}

\bibitem{BZ16}
{O. Bello and S. Zeasally}, ``Intelligent device-to-device communication in the
  internet of things,'' {\em IEEE Systems Journal}, vol.~10, pp.~1172--1182,
  September 2016.

\bibitem{GJ16}
{P. Gandotra and R.K. Jha}, ``Device-to-device communication in cellular
  networks: A survey,'' {\em Elsevier Journal of Network and Computer
  Applications}, vol.~71, pp.~99--117, August 2016.

\bibitem{TUY14}
{M.N. Tehrani and M. Uysal and H. Yanikomeroglu}, ``Device-to-device
  communication in {5G} cellular networks: {C}hallenges, solutions, and future
  directions,'' {\em IEEE Communications Magazine}, vol.~52, pp.~86--92, May
  2014.

\bibitem{CGS13}
{D. Camps-Mur, A. Garcia-Saavedra and P. Serrano}, ``Device-to-device
  communications with {Wi-Fi Direct}: {Overview} and experimentation,'' {\em
  IEEE Wireless Communications}, vol.~20, pp.~96--104, June 2013.

\bibitem{FDM12}
{G. Fodor and E. Dahlman and G. Mildh and S. Parkvall and N. Reider and G.
  Miklos and Z. Turanyi}, ``Design aspects of network assisted device-to-device
  communications,'' {\em IEEE Communications Magazine}, vol.~50, pp.~170--177,
  March 2012.

\bibitem{Detti2015}
{A. Detti , B. Ricci and N. Blefari-Melazzi}, ``Mobile peer-to-peer video
  streaming over information-centric networks,'' {\em Elsevier Computer
  Networks}, vol.~81, pp.~272--288, April 2015.

\bibitem{LJK15}
K.-W. Lim, W.-S. Jung, and Y.-B. Ko, ``Energy efficient quality-of-service for
  wlan-based d2d communications,'' {\em Ad Hoc Networks}, vol.~25,
  pp.~102--116, February 2015.

\bibitem{Mumtaza2014}
{S. Mumtaza, H. Lundqvistb, K. M. Saidul Huqa, J. Rodrigueza and A. Radwan},
  ``Smart {Direct-LTE} communication: {An} energy saving perspective,'' {\em
  Elsevier Ad Hoc Networks}, vol.~13, pp.~296--311, February 2014.

\bibitem{Jahed2012}
{K. Jahed, M. Younes and S. Sharafeddine}, ``Energy measurements for mobile
  cooperative video streaming,'' {\em IFIP Wireless Days}, November 2012.

\bibitem{SRM16}
V.~Sucasas, A.~Radwan, H.~Marques, J.~Rodriguez, S.~Vahid, and R.~Tafazolli,
  ``A survey on clustering techniques for cooperative wireless networks,'' {\em
  Ad Hoc Networks}, vol.~47, pp.~53--81, September 2016.

\bibitem{ACF17}
M.~A. Khan, W.~Cherif, F.~Filali, and R.~Hamila, ``Wi-fi direct research -
  current status and future perspectives,'' {\em Journal of Network and
  Computer Applications}, vol.~93, pp.~245--258, September 2017.

\bibitem{Haider2015}
{N. Haider, M. Imran, M. Younis, N. Saad and M. Guizani}, ``A novel mechanism
  for restoring actor connected coverage in wireless sensor and actor
  networks,'' {\em IEEE ICC 2015 - Ad hoc and Sensor Networking Symposium},
  June 2015.

\bibitem{Younisa2014}
{M. Younisa, I. F. Senturkb, K. Akkayab, S. Leec and F. Seneld}, ``Topology
  management techniques for tolerating node failures in wireless sensor
  networks: A survey,'' {\em Elsevier Computer Networks}, vol.~58,
  pp.~254--283, January 2014.

\bibitem{Kofahi2010}
{O. M. Al-Kofahi and A. E. Kamal}, ``Survivability strategies in multihop
  wireless networks,'' {\em IEEE Wireless Communications}, vol.~17, pp.~71--80,
  October 2010.

\bibitem{GZH12}
M.~Garrosi, B.~Zafar, and M.~Haardt, ``Prolonged network life-time in
  self-organizing peer-to-peer networks with e-rssi clustering,'' in {\em IEEE
  International Conference on Communications (IEEE ICC)}, June 2012.

\bibitem{SY15}
A.~Shahin and M.~Younis, ``Efficient multi-group formation and communication
  protocol for wi-fi direct,'' in {\em Conference on Local Computer Networks
  (LCN)}, October 2015.

\bibitem{AKJ15}
M.~Azharuddin, P.~Kuila, and P.~Jana, ``Energy efficient fault tolerant
  clustering and routing algorithms for wireless sensor networks,'' {\em
  Computers \& Electrical Engineering}, vol.~41, pp.~177--190, January 2015.

\bibitem{SJF17}
S.~Sharafeddine, K.~Jahed, O.~Farhat, and Z.~Dawy, ``Failure recovery in
  wireless content distribution networks with device-to-device cooperation,''
  {\em Computer Networks}, April 2017.

\bibitem{CYF15}
P.~Chaki, M.~Yasuda, and N.~Fujita, ``Seamless group reformation in wifi peer
  to peer network using dormant backend links,'' in {\em 12th Annual IEEE
  Consumer Communications and Networking Conference (CCNC)}, January 2015.

\bibitem{PJ17}
S.~Pathak and S.~Jain, ``An optimized stable clustering algorithm for mobile ad
  hoc networks,'' {\em EURASIP Journal on Wireless Communications and
  Networking}, vol.~51, pp.~1--11, March 2017.

\bibitem{MMZ15}
N.~Mansoor, A.~M. Islam, M.~Zareei, S.~Baharun, and S.~Komaki, ``Construction
  of a robust clustering algorithm for cognitive radio ad-hoc network,'' in
  {\em International Conference on Cognitive Radio Oriented Wireless Networks
  (CrownCom 2015)}, April 2015.

\bibitem{SRV14}
T.~{Shiva Prakash}, K.~Raja, K.~Venugopal, S.~Iyengar, and L.~Patnaik, ``Fault
  tolerant qos adaptive clustering for wireless sensor networks,'' in {\em
  Ninth International Conference on Wireless Communication and Sensor
  Networks}, 2014.

\bibitem{JZZ16}
D.~Jia, H.~Zu, S.~Zou, and P.~Hu, ``Dynamic cluster head selection method for
  wireless sensor network,'' {\em IEEE Sensors Journal}, vol.~16,
  pp.~2746--2754, April 2016.

\bibitem{ND16}
P.~Nayak and A.~Devulapalli, ``A fuzzy logic-based clustering algorithm for wsn
  to extend the network lifetime,'' {\em IEEE Sensors Journal}, vol.~16,
  pp.~137--144, January 2016.

\bibitem{HCB02}
W.~Heinzelman, A.~Chandrakasan, and H.~Balakrishnan, ``An application-specific
  protocol architecture for wireless microsensor networks,'' {\em IEEE
  Transactions on Wireless Communications}, vol.~1, pp.~660--670, October 2002.

\bibitem{CKF17}
W.~Cherif, M.~Khan, F.~Filali, S.~Sharafeddine, and Z.~Dawy, ``P2p group
  formation enhancement for opportunistic networks with wi-fi direct,'' in {\em
  IEEE Wireless Communications and Networking Conference (IEEE WCNC)}, March
  2017.

\bibitem{JFA16}
K.~Jahed, O.~Farhat, G.~Al-Jurdi, and S.~Sharafeddine, ``Optimized group owner
  selection in wifi direct networks,'' in {\em 24th International Conference on
  Software, Telecommunications and Computer Networks (SoftCOM)}, September
  2016.

\bibitem{MCB14}
U.~Menegato, L.~Cimino, S.~Delabrida, F.~Medeiros, R.~Oliveira, and D.~Ufop,
  ``Dynamic clustering in wifi direct technology,'' in {\em 12th ACM
  International Symposium on Mobility Management and Wireless Access (MobiWac
  2014)}, September 2014.

\bibitem{LSY16}
K.~Liu, W.~Shen, B.~Yin, X.~Cao, L.~Cai, and Y.~Cheng, ``Development of mobile
  ad-hoc networks over wi-fi direct with off-the-shelf android phones,'' in
  {\em IEEE International Conference on Communications (ICC)}, May 2016.

\bibitem{JSM16}
{K. Jahed and S. Sharafeddine and A. Moussawi and A. Abou Daya and H. Dbouk and
  S. Kassir and Z. Dawy and P. Valsalan and W. Cherif and F. Filali},
  ``Scalable multimedia streaming in wireless networks with device-to-device
  cooperation,'' {\em ACM Multimedia 2016 Conference}, October 2016.

\bibitem{balasubramanian2009energy}
N.~Balasubramanian, A.~Balasubramanian, and A.~Venkataramani, ``Energy
  consumption in mobile phones: a measurement study and implications for
  network applications,'' in {\em Proceedings of the 9th ACM SIGCOMM Conference
  on Internet Measurement}, pp.~280--293, ACM, 2009.

\bibitem{6952687}
N.~Abbas, Z.~Dawy, H.~Hajj, and S.~Sharafeddine, ``Energy-throughput tradeoffs
  in cellular/wifi heterogeneous networks with traffic splitting,'' in {\em
  2014 IEEE Wireless Communications and Networking Conference (WCNC)},
  pp.~2294--2299, April 2014.

\bibitem{Christos0210040}
C.~H. Papadimitriou, ``Worst-case and probabilistic analysis of a geometric
  location problem,'' {\em SIAM Journal on Computing}, vol.~10, no.~3,
  pp.~542--557, 1981.

\bibitem{Nimrod0213014}
N.~Megiddo and K.~Supowit, ``On the complexity of some common geometric
  location problems,'' {\em SIAM Journal on Computing}, vol.~13, no.~1,
  pp.~182--196, 1984.

\bibitem{Force}
M.~Khandani, P.~Saeedi, Y.~Fallah, and M.~Khandani, ``A novel data clustering
  algorithm based on electrostatic field concepts,'' in {\em IEEE Symposium on
  Computational Intelligence and Data Mining (CIDM)}, March 2009.

\bibitem{Force1}
M.~Khandani, R.~Bajcsy, and Y.~Fallah, ``Automated segmentation of brain tumors
  in mri using force data clustering algorithm,'' in {\em International
  Symposium on Visual Computing}, November 2009.

\bibitem{4031363}
M.~X. Goemans, ``Minimum bounded degree spanning trees,'' in {\em 47th Annual
  IEEE Symposium on Foundations of Computer Science (FOCS'06)}, October 2006.

\bibitem{lloyd}
S.~Lloyd, ``Least squares quantization in pcm,'' {\em IEEE Transactions on
  Information Theory}, vol.~28, pp.~129--137, September 2006.

\bibitem{Halperin:2010:DPC:1924920.1924928}
D.~Halperin, B.~Greenstein, A.~Sheth, and D.~Wetherall, ``Demystifying 802.11n
  power consumption,'' in {\em International Conference on Power Aware
  Computing and Systems}, 2010.

\end{thebibliography}

\end{document}